\documentclass[%
 aip,
 jmp,%
 amsmath,amssymb,
preprint,nofootinbib%
]{revtex4-1}

\usepackage{graphicx}
\usepackage{dcolumn}
\usepackage{bm}
\usepackage{amsfonts}
\usepackage{color}

\usepackage{ccaption,subfig}

\usepackage[export]{adjustbox}
\usepackage{floatrow}

\usepackage{float}
\floatstyle{plaintop}
\restylefloat{table}

\usepackage{footmisc}
{
\makeatletter
\def\frontmatter@thefootnote{%
 \altaffilletter@sw{\@fnsymbol}{\@fnsymbol}{\csname c@\@mpfn\endcsname}%
}}%

\makeatother

\begin{document}

\title{Topological transitions, turbulent-like motion and long-time-tails driven by cell division in biological tissues }


\author{Xin Li}
\affiliation{Department of Chemistry, University of Texas, Austin, TX 78712, USA}
\author{Sumit Sinha}
\affiliation{Department of Physics, University of Texas, Austin, TX 78712, USA}
\author{ T. R. Kirkpatrick}
\affiliation{lnstitute of Physical Science and Technology, University of Maryland, College Park, Maryland 20742, USA}
\author{D. Thirumalai}
\email{dave.thirumalai@gmail.com}
\affiliation{Department of Chemistry and Department of Physics, University of Texas, Austin, TX 78712, USA}

\date{\today}

\begin{abstract}
The complex spatiotemporal flow patterns in living tissues, driven by active forces, have many of the characteristics associated with inertial turbulence even though the Reynolds number is extremely low. Analyses of experimental data from two-dimensional epithelial monolayers in combination with agent-based simulations show that cell division and apoptosis lead to  directed cell motion for hours, resulting in rapid topological transitions  in neighboring cells. These transitions in turn generate both long ranged and long lived clockwise and anticlockwise vortices, which gives rise to turbulent-like flows. Both experiments and simulations show that at long wavelengths the  wave vector ($k$) dependent energy spectrum $E(k) \approx k^{-5/3}$, coinciding with the Kolmogorov scaling in fully developed inertial turbulence. Using theoretical arguments and simulations, we show that long-lived vortices lead to long-time tails in the velocity auto-correlation function, $C_v(t) \sim t^{-1/2}$, which has the same structure as in classical 2D fluids but with a different scaling exponent.
\end{abstract}


\pacs{}

\maketitle

\section{Introduction}
The cell, an active agent, is the basic unit of life. The motion of a single cell has been studied intensively in the past few decades\cite{lauffenburger1996cell,horwitz1999cell,bray2000cell,ponti2004two,li2009actin,raab2012crawling,novikova2017persistence,van2018mechanoreciprocity}. However, cells interact with each other and move in a collective manner, which plays an essential role in many biological processes, such as embryonic morphogenesis, organ development, wound healing, and spreading of cancer\cite{friedl2009collective,rorth2009collective,poujade2007collective}.  The collective movement of cells using physical models, spanning from sub-cellular to supra-cellular length scales\cite{hakim2017collective,camley2017physical,alert2020physical}, gives rise to unexpected spatial and temporal correlations.


Since the discovery that flow-induced stretching of a long polymer in a high viscosity solution could produce flow patterns resembling that found in fully developed turbulence~{\cite{groisman2000elastic}}, a number of studies have shown that turbulent-like features  emerge in a variety of biological systems \cite{alert2022active,dombrowski2004self,rossen2014long,bratanov2015new,wensink2012meso}. In these examples, the complex flow patterns are thought to rise from active forces, and hence the collective behavior is referred to as active turbulence, even though the scaling behavior is not universal. Hydrodynamic models, that build on the Toner-Tu equations \cite{toner1998flocks,toner2005hydrodynamics,wensink2012meso,bratanov2015new} and particle based simulations for self-propelled rods \cite{wensink2012meso} have generated turbulent-like patterns. In addition, numerical solutions and theoretical arguments of active nematic liquid crystal models have been used to show that vortex formation and turbulent-like behavior emerges \cite{giomi2015geometry,thampi2015intrinsic,alert2020universal}. In all these models, active forces, the cause for turbulent-like behavior, is introduced phenomenologically (by ``hand" as it were).

Recent studies\cite{rossen2014long,doostmohammadi2015celebrating} suggest that cell division and apoptosis lead to turbulent-like velocity fields with long-range vortex structures in two dimensional (2D) cell monolayers upon averaging over hundreds of division events. In addition, it is found that cell displacements exhibit super-diffusive behavior in a 2D confluent monolayer \cite{giavazzi2018tracking}, where cell division and apoptosis occur continuously. By building  on our previous studies\cite{malmi2018cell,sinha2020spatially}, which showed that imbalance between the rates of cell division and apoptosis, leads to unusual dynamics  in a growing tumor spheroid, we investigate if a similar picture could be used to analyze the time-dependent cell trajectories generated  in the 2D monolayer\cite{giavazzi2018tracking} to generate new physical properties. 

Here, we investigate the collective cell migration induced by division and apoptosis in a confluent 2D cell monolayer using experimental data on Madin-Darby Canine Kidney (MDCK) cells\cite{giavazzi2018tracking}, in combination with an agent-based simulations  \cite{malmi2018cell,sinha2020spatially}. We find that cell division leads to a directed motion of cells, which explains the  super-diffusive behavior of cells with the mean-square displacement, $\Delta(t) \propto t^{1.5}$.  The directed motion of the new born cells induce rapid rearrangements of the neighbors in the confluent tissue through topological (T1 and T2 transitions) changes.  The cell rearrangement induced by a single cell division leads to vortex formation.   With the occurrence of multiple cell births and deaths, a turbulent-like behavior naturally emerges with the mean size of the vortices being $\sim$ 100 $\mu m$.  Interestingly, we find two scaling regimes for the kinetic energy spectrum, $E(k)$ with  one  following a Kolmogorov-like power law ($E(k) \approx k^{-5/3}$). The long-range correlated vortex motion leads to a long time tail (LTT)  for the velocity auto-correlation function ($C_v(t)$), which decays as $C_v(t) \sim t^{-1/2}$, which  differs from the well-known result $C_v(t) \sim \frac{1}{t \sqrt{ln t}}$ found in classical hard-disc-model fluids. 

 
\section{Results}


\textbf{Vortex motion in a confluent cell monolayer:}
A confluent MDCK monolayer exhibits rich dynamics (see Figs.~S1-S5 for some general features of the monolayer as it evolves in time) even though the cells are jammed.  To probe the motility of the cells in the collective, we created the velocity vector field  by applying the digital particle image velocimetry\cite{thielicke2014pivlab} to the experimental data\cite{giavazzi2018tracking}.   Collective motion of cells, in which a group of cells move in the same direction with similar magnitude of velocity (${\bf v}$)\cite{vicsek2012collective}, occurs in different regions of the monolayer (see Fig.~\ref{figmodel}a). Interestingly, formation of several vortices are discernible in the velocity field, which indicates a coherent motion of cells (see the dashed rectangle in Fig.~\ref{figmodel}a as an example).  We calculated the vorticity field (${\boldsymbol {\omega}} \equiv  \nabla \times {\bf v} $), which describes the local rotational motion of the velocity flow (see Fig.~\ref{figmodel}b).  Positive vorticity (yellow color) indicates clockwise rotation, while negative vorticity (blue color) shows anticlockwise rotation.  The vortices in  Fig.~\ref{figmodel}a are clearly elucidated  in the vorticity field (see the bright yellow circle in the lower left of Fig.~\ref{figmodel}b and the dashed rectangle in Fig.~\ref{figmodel}a).  Outward and inward velocity fluxes, characterized by divergence ($\text{div} ({\bf v}) \equiv  \nabla \cdot {\bf v} $) of the velocity flow (Fig.~\ref{figmodel}c),  are non-uniform in the monolayer.  



\textbf{Cell division as the driver of vortex motion:}  Previous theoretical and experimental studies\cite{ranft2010fluidization,matoz2017cell,malmi2018cell,petridou2019fluidization} have shown that   active forces generated by cell division  drives fluidization of tissues, leading to heterogeneous cell motility patterns both in a two-dimensional monolayer and in the three-dimensional spheroids. Such events (Fig.~S1c) occur frequently in the experiments, as shown in Figs.~\ref{figmodel}a-c.  Therefore, we surmised that the complex flow patterns observed here are driven by cell birth and death.  To test this notion,  we first show the positions of newly generated cells (born about 30 minutes before the snapshot in Fig.~\ref{figmodel}a) in the divergence field (see the purple dots in Fig.~\ref{figmodel}c).  The origin of the outward flux (colored in yellow) is frequently accompanied by the appearance of newly created cells nearby.  In addition, new cells arising from cell divisions are  close to the vortices (see Figs.~\ref{figmodel}b-c).  Therefore, it is likely that the flow patterns (Figs.~\ref{figmodel}a-c) in the monolayer are driven by  active forces produced by the cell division.

\textbf{Particle based simulations:} To further elucidate the relevance of cell division, we simulated the dynamics of a cell monolayer using an agent-based model (details are in the Methods section and the SI). The velocity flow pattern in the simulations and experiments are similar (compare Figs.~\ref{figmodel}a and \ref{figmodel}d).  In addition, several pairs of vortices with opposite direction of rotation are also observed (see the lower middle of Fig.~\ref{figmodel}e). Note that in contrast to other studies\cite{belmonte2008self,basan2013alignment,bi2016motility}, we do not include self-propulsion in our model. Thus, the observed velocity flow and vortex formation can only arise from self-generated active forces (SGAFs) arising from cell division.  Indeed, cell division (see the purple dots in Fig.~\ref{figmodel}f), which produces an outward flow in the the divergence field (Fig.~\ref{figmodel}f),  is close to the vortex location (see Figs.~\ref{figmodel}d-e).  Hence, cell division and apoptosis generate complex cell motility patterns, resembling turbulent-like structures (Figs.~\ref{figmodel}a-f).

To illustrate how cell division leads to vortex formation, we tracked the collective cell motion in the dashed rectangle region in Fig.~\ref{figmodel}a for one hour (see Fig.~\ref{figmodel}g-l). The mother cell (in pink, see Fig.~\ref{figmodel}g at $t =100$ min) divides into two daughter cells (orange and blue) at $t = 110$ min (see Fig.~\ref{figmodel}h). The vortex observed in Fig.~\ref{figmodel}a-b at $t = 120$ min is established at $t = 110$ min (see both the velocity and vorticity fields in Fig.~\ref{figmodel}h) immediately after  cell division. The vortex persists for an additional 20 minutes (Fig.~\ref{figmodel}i-j) before vanishing (Fig.~\ref{figmodel}k-l). In contrast to previous experimental observations\cite{rossen2014long,doostmohammadi2015celebrating}, where vortices are discernible only upon averaging over 100 division events, we found that a single cell division event leads to a vortex,  lasting for more than 30 minutes. In addition, multiple cell division events drive a turbulent-like flow (Fig.~\ref{figmodel}a-f).



\textbf{Directed motion of new born cells leads to anomalous diffusion:}
To study how vortices are formed by cell division explicitly, we started by monitoring  the cell dynamics in the trajectories during the  first $100$ minutes for the newly created cells  (see Fig.~\ref{msddis}a).  For illustration purposes, we set the initial positions of all the cells at the origin. There is no obvious preferred direction of motion for cells, which suggests that they move isotropically (Fig.~\ref{msddis}a). Surprisingly, the motion is directed after cell division  over a distance of  $\sim$ 10 $\mu m$ (see Fig.~\ref{msddis}a and the inset figure for a few representative trajectories).  The simulations also show a similar directed motion of cells during the first 100 minutes after cell divisions (see Fig.~\ref{msddis}b). 


To illustrate the consequence of cell-division induced directed motion, we calculated the  mean-square displacement ($\Delta(t, t_i) =\langle [\textbf{r}(t_i+t)-\textbf{r}(t_i)]^2 \rangle$) of cells at different initial times $t_i$, where \textbf{r} is the cell position. Both experiments and simulations (Figs.~\ref{msddis}c-d) exhibit  superdiffusive behavior, $\Delta(t, t_i) \propto t^\alpha$, with an exponent $\alpha \approx 1.51 \pm 0.1$ and $1.44 \pm 0.1$ respectively, irrespective of the time windows used.   The displacement distributions, ($P(\Delta x)$, $P(\Delta y)$), of cells calculated from both experiments and simulations (see Fig.~S6a-d in the SI) deviate from the Gaussian distribution (see the green solid lines). It follows that cell division and apoptosis is the origin of superdiffusive motion of cells.


\textbf{Cell division and apoptosis result in topological changes:}
The SGAFs \cite{sinha2020self} due to cell division and apoptosis lead to a directed motion of newly generated cells,  which  influences the movement of cells nearby in a confluent tissue. Interestingly, there are several T1 transitions, which are critical during morphogenesis\cite{irvine1994cell}, and cancer invasion\cite{levayer2015cell},  during a single cell division in the MDCK cell experiments. One example is shown in Figs.~\ref{msddis}e-g (see also Fig.~S7 in the SI). The force produced by the division of cell $1$ at $t = 110$ min separates the two neighboring cells (cells 2 and 4) away from each other but brings cells $1$ and $3$ together at later times.  Such rearrangement of cells $1-4$ and other cells nearby finally leads to vortex formation (see the velocity vector field in Figs.~\ref{msddis}e-g and also Figs.~\ref{figmodel}g-l). 



In addition to a single vortex induced by one cell division, (Figs.~1g-l), there is a pair of vortices,  with opposite sense of rotation,  in certain time frames (see the dashed rectangle in Fig.~\ref{clock}a). In the same region, there are three cell divisions, which lead to dramatic cell rearrangements nearby, 10 minutes before a vortex pair formation (the purple dots in the the dashed rectangle). To describe the formation of one vortex pair quantitatively, we plot the vorticity value along the arrow in the dashed rectangle in Fig.~\ref{clock}a using the digital particle image velocimetry\cite{thielicke2014pivlab}. A smooth transition from anticlockwise to clockwise vorticity is clearly depicted (see Fig.~\ref{clock}b).  Similarly, several pairs of vortices with opposite directions of rotation are  found in simulations (see the lower middle of Fig.~\ref{figmodel}e).  Besides, there is evidence for T2 transition (Fig.~S8 in the SI), which arises from the extrusion (apoptosis) of a cell from the confluent monolayer. Taken together, these results suggest that cell division and apoptosis not only regulate cell numbers during morphogenesis and cancer development, they also drive dramatic cell rearrangements, as discovered recently for chick gastrulation\cite{firmino2016cell}, and germband extension of Drosophila\cite{da2007oriented}. Finally, the cell rearrangement through topological T1/T2 transitions leads to the formation of vortices and turbulent-like structure in the confluent MDCK tissue.

\textbf{Spatial correlations:}  To probe the spatial variations in the turbulent-like flow, we calculated space-dependent correlations associated using the velocity field, $C_{v}(R)$. From the cell position, $\textbf{r}_{i}(t)$, of the $i^{th}$ cell at different times, we calculated the velocity, $\textbf{v}_{i}(t) \equiv [\textbf{r}_{i}(t+\delta t/2)-\textbf{r}_{i}(t-\delta t/2)]/{\delta t}$, at time $t$ where $\delta t$ is the time interval. The spatial velocity correlation function ($C_{v}(R)$) is defined as,
\begin{equation}
C_{v}(R) = \langle \frac{\sum_{i}(\bf{v}(r_{i})-\langle v(r_{i})\rangle)\cdot(v(r_{i}+R)-\langle v(r_{i})\rangle)}{\sum_{i}(\bf{v}(r_{i})-\langle v(r_{i})\rangle) \cdot (v(r_{i})-\langle v(r_{i})\rangle)} \rangle .
\label{cdr}
\end{equation}
The correlation function at different times (see Fig.~\ref{cvt}a) has negative minimum at a distance $R_{v} \approx 100~\mu m$, which is about $10$ times greater than the mean cell size (see Fig.~S3). The negative values for the correlation function are due to the anti-parallel velocity on the opposite sides of the vortices (see Fig.~\ref{figmodel}b). Therefore, $R_{v}$ is roughly the size of the vortices (Fig.~\ref{figmodel}), which are about 100 $\mu m$. The $C_{v}(R)$ calculated from our simulations  supports such a  link (see Fig.~\ref{cvt}b). Interestingly, the collective motion of cells is quite robust, persisting for a long time, despite the increase in the cell density (Fig.~S4a).

\textbf{Long time tails (LTT):}  We calculated the velocity correlation function ($C^{i}_{v}(t)$) for cell $i$,  
\begin{equation}
C^{i}_{v}(t) = \langle \frac{[\textbf{v}_{i}(t_{0}+t)- \langle\textbf{v}_{i} \rangle]\cdot [\textbf{v}_{i}(t_{0})- \langle\textbf{v}_{i} \rangle]}{[\textbf{v}_{i}(t_{0})- \langle\textbf{v}_{i} \rangle]\cdot [\textbf{v}_{i}(t_{0}) -\langle\textbf{v}_{i} \rangle]} \rangle ,
\end{equation}
where $\langle\textbf{v}_{i} \rangle$ is the mean velocity of cell $i$, and the average is over time $t_{0}$ (24 hours).  The ensemble average $C_{v}(t)$ for the  monolayer is obtained by averaging $C^{i}_{v}(t)$ over all the cells.  The decay of $C_{v}(t)$ follows  $C_v(t) \sim t^{-\beta}$  with  $\beta = 0.4 \pm 0.1$ (see the blue dash-dotted line in Fig.~\ref{cvt}c).  Our simulations also show a power-law decay with  $\beta = 0.59 \pm 0.03$ (see Fig.~\ref{cvt}d). Based on dimensional argument, we expect that the decay of  $C_{v}(t) \sim \Delta(t)/t^2 \sim t^{-0.5}$, which is fairly close to the experimental and simulation results. Using theory that accounts for SGAFs, the exponent $\beta = 1/2$ in 2D (see the  Methods section).  A comparison from three different fit functions to the experimental and simulation data are also shown in the Fig.~S9 in the SI.  The emergence of LTT in $C_{v}(t)$ shows  that motion of cells is persistent in a specific direction (Figs.~\ref{msddis}a-d), which in turn is linked to vortex motion.   To ascertain whether the power-law relation depends on the time interval selected to calculate $C_{v}(t)$, we varied $\delta t$. The  value of $\beta$ stays around 1/2 (see Fig.~S10 in the SI). 

A long time tail relation is also found for the  current-current correlation functions in disordered systems, which decays by $t^{-d/2}$ ( $d$ is the dimension of  space)\cite{belitz1994anderson}. Such a relation was first reported in simulations of  classical hard-disc (sphere)-model fluids\cite{alder1970decay}, and was explained by correlated collision events  theoretically\cite{ernst1970asymptotic,dorfman1970velocity}. 
The difference in the value of $\beta$ between abiotic systems and MDCK monolayer is due to the generation  of SGAFs due to cell division, which results in persistent (almost ballistic) directional motion of cells for long times.



\textbf{Energy spectrum:} It is clear that turbulent-like flow emerges naturally in the collective motion of cells (Figs.~\ref{figmodel}a-c).  To characterize the nature of turbulent-like motion in the epithelial cells, we calculated the wave vector ($k$) dependent energy spectrum, $E(k)$, at different times (see Fig.~\ref{cvt}e). There are two scaling regimes in  $E(k)$ as a function of $k$. In the intermediate values of $k$,  $E(k)\sim k^{-1.58(0.2)}$. The value of the exponent is close to the Kolmogorov-Kraichnan prediction, $-5/3$, found in the inertial turbulence\cite{kolmogorov1941local}. We found a similar exponent value ($-1.4\pm0.2$) from our simulations (see Fig.~\ref{cvt}f).  In the MDCK cells,  the Reynolds number is small ($\approx 0$)\cite{marchetti2013hydrodynamics}, which shows that the underlying mechanisms must be quite different for the turbulent-like motion\cite{alert2022active}. 

At smaller scales (large $k$), the results (Fig.~4e) obtained by analyzing the experimental trajectories show that $E(k)\sim k^{-\lambda}$ ($\lambda = 3.5 \pm 0.2$). From the agent-based simulations, we deduce that $E(k)\sim k^{-3.7(0.2)}$. Two comments are worth making: $(i)$ The  $\lambda$ values from experiments and simulations are fairly close. It is interesting that scaling anticipated in the context of inertial turbulence is obeyed in living systems, given that energy injection, which is a consequence of cell division and apoptosis, is autonomous.  $(ii)$ The prediction for inertial 2D turbulence at large $k$ is $\lambda \approx 3$, which differs from the estimated value for MDCK cells and simulations. It is unclear if the large $k$ finding in living tissues is simply a consequence of non-universal behavior or if large $k$ limit is not accessed in experiments and simulations. In active systems, such as microtubule-kinesin complex and bacterial suspensions, $\lambda$ ranges from 2 to 4.5\cite{wensink2012meso,lin2021energetics,guillamat2017taming,alert2022active}.
 

\section{Discussion:}
We investigated the dynamics of cell rearrangement and collective motion in a biologically relevant tissue \cite{friedl2009collective,rorth2009collective,poujade2007collective}, by analyzing data from  confluent MDCK cell monolayers, combined with  agent-based simulations and theory. Cell division and apoptosis lead to  a directed motion of cells for hours, which results in a universal super-diffusive behavior of cells characterized by the  mean-square displacement exponent,  $\alpha \approx 1.51$, irrespective of the initial time $t_i$ considered in the experiments\cite{giavazzi2018tracking}. Our agent-based simulations  yield $\alpha = 1.44\pm 0.1$. It is worth emphasizing that, in the absence of cell division, there is complete cessation of motion in the simulations. Therefore, the complex dynamics in the simulations, which capture the salient features in the MDCK cells, arises solely due to self-generated active forces generated by cell division and apoptosis\cite{malmi2018cell,sinha2020spatially,sinha2020self}.

The directed motion of newly generated cells induces rapid rearrangements of cells nearby through T1--cell intercalation and T2 transitions. Interestingly, cell rearrangements due to a single cell division result in vortex formation in confluent tissues, with lifetimes lasting for 30 minutes.  More importantly, turbulent-like flux patterns emerges in the region where multiple divisions occur, which is also supported by our agent-based simulation model. From the  spatial velocity correlation function, we estimate that the mean size of the vortices  can be as large as to 100 $\mu m$, which is nearly 10 times larger than the size of a single cell ($\sim 10~\mu m$). Finally, the kinetic energy spectrum ($E(k)$) exhibits two distinct scaling behaviors of the collective cell movement. At intermediate values of $k$, $E(k) \sim k^{1.58\pm 0.2}$, which is close to the predicted Kolmogorov-Kraichnan behavior for  the inertial turbulence. At large $k$ value, $E(k) \sim k^{-3.5}$, which does not seems to have counter part in inertial turbulence, and is likely to be non-universal. The exponent characterizing large $k$ behavior of $E(k)$ show substantial variations, depending on the system\cite{alert2022active,blanch2018turbulent,mueller2019emergence}.


In classical hard-disc-model fluids\cite{alder1970decay,ernst1970asymptotic,dorfman1970velocity}, a long time tail is expected for the velocity autocorrelation function with $C_{v}(t) \propto \frac{1}{t \sqrt{ln t}}$.  Both experiments and simulations show a slower power law decay  $C_{v}(t)\propto t^{-1/2} $ in the MDCK cell monolayer.    To explain this finding, we developed a theory that accounts for active forces leading to the prediction of a slower decay,   $C_{v}(t) \propto t^{-1/2}$.  The theory and the agent-based simulations indicate that self-generated active forces lead to highly correlated responses. Given the simple model considered here, we expect that our studies can be used to describe collective motion of other cell types or a mixture of different cell types, and even other active systems in general.

In addition, we showed that the amplitude for the mean-square displacement (Fig.~2c) and the mean velocity of cells (Fig.~S4b) decrease, as the initial time $t_i$ increases, which indicates a slowing down due to jamming or aging of the cell dynamics. Similar aging behavior has been noted in other types of cell monolayers\cite{garcia2015physics}. In addition, all the dynamical characteristics for cells,  (superdiffusion, compressed exponential relaxation and aging) are found in many soft glassy systems\cite{bouzid2017elastically,gnan2019microscopic,galloway2020scaling} in the absence of active forces. Thus the physics used to describe abiotic glassy materials could be applied to understand living active matter\cite{kirkpatrick2015colloquium,tjhung2020analogies}. Finally, more detailed models by incorporating the size reductive divisions\cite{puliafito2012collective}, maturation and strengthening of cell-cell adhesion\cite{garcia2015physics} or other potential processes could be developed based on our present study to explain aging and other complex cell activities.

\section{Methods:}
\textbf{Summary of  MDCK epithelial cell experiments\cite{giavazzi2018tracking}:} The MDCK (Madin-Darby Canine Kidney) epithelial cells were maintained in Dulbeccos Modified Eagle Medium, supplemented with $5\%$ fetal bovine serum and $1\%$ L-Glutamine. First, they were seeded in a 6-well plate to grow in complete medium until a confluent cell monolayer  formed. Then, the images of the cell monolayer were acquired by Olympus IX81 inverted microscope (both in phase contrast and wide-field fluorescence) every minute for a 24-hour period. The phase-contrast microscopy was used to visualize the cell membrane and different organelles (see Fig.~S1a). The nuclei of the cells at each time frame were also captured by wide-field fluorescence microscopy of EGFP-H2B expressing cells (see the cell nuclei shown in Fig.~S1b for an example) and their positions were recorded from the single particle tracking by using the ImageJ\cite{rueden2017imagej2}.    A number of cell division events were observed in such a confluent monolayer (see Fig.~S1c), while cell apoptosis/extrusion process led to the removal of cells from the monolayer. From the cell nuclei positions at each time frame, the cell trajectories can be obtained (see Fig.~S1d,  shows the first 150 minutes of recording for one of the field-of-views (FOV)). The cell trajectory in the figure exhibits a highly heterogeneous behavior. Some cells move in a rather straight and regular trajectory, while others exhibit irregular curly motion. The diverse movement is even clearer if the cell trajectories are plotted over the whole 24-hour period  (see the two examples shown in the insets of Fig.~S1d). The dynamically heterogeneous behavior is  reminiscent of supercooled liquids, as was first shown by Angelini et al.\cite{angelini2011glass} in the context of cell monolayers.


\textbf{Particle-based simulations:}
We used a two-dimensional version of a particle-based model \cite{malmi2018cell,sinha2020spatially} to simulate cell dynamics. We mainly focused on investigating the roles of the cell division and apoptosis on the collective cell motion in order to provide microscopic insights into experiments\cite{giavazzi2018tracking}. A single cell is modeled as a two-dimensional deformable disk. We use a periodic boundary conditions to mimic large systems. The cells interact with each other through repulsive (Hertzian contact mechanics)  and attractive interactions (cell-cell adhesion). The motion of cells is described by an overdamped Langevin equation. The cells were allowed to grow and divide, as in the experiment. We also removed the cells from the monolayer at a constant rate to model the cell apoptosis and extrusion events that were observed in experiments. Because we do not consider self-propulsion, the cell motility can only arise  from the self-generated active forces  through cell growth, division and apoptosis\cite{sinha2020self}.  The SI contains details of the model.

\textbf{Theory of LTT driven by cell division:}
The turbulent-like motion in the MDCK cells in which active forces are generated by cell division, suggests that we use methods developed to treat turbulence in conventional fluids \cite{forster1977large,dedominicis1979energy,yakhot1986renormalization,smith1992yakhot}. The fundamental difference is that  the random stirring that causes the turbulence in the conventional case is replaced by the internally generated cell division process in the biological case. Our aim is to compute the turbulent renormalized transport coefficients in the biological case, and from that determine the diffusive velocity autocorrelation function and the mean-squared-displacement. The relevant spatial dimension is $d=2$.

The model dynamical equations are,
\begin{equation}
\partial_tu_i+u_j\nabla_ju_i=-\nabla_ip+\nu_0\nabla^2u_i+f_i
\label{veloeq}
\end{equation}
\begin{equation}
\partial_tc+u_j\nabla_jc=D_0\nabla^2c+g
\end{equation}
\begin{equation}
\nabla\cdot\mathbf{u}=0
\label{ueq}
\end{equation}
Here, $u_i$ is the fluid velocity in the $i$-direction (repeated indices are to be summed over), $p$ is the pressure, $\nu_0$ is the bare kinematic viscosity, $c$ is the density of the active matter (MDCK cells), $D_0$ is a bare diffusion coefficient. The equations given above couple vorticity in the fluid motion to cell density. We have also imposed the incompressibility condition in Eq.~(\ref{ueq}). The statistical forces $f_i$ and $g$ are Gaussian noise terms that have zero mean and a nonzero two-point correlation functions that are local in space and time  with gradient independent correlation strengths given by $\Delta_f$ and $\Delta_g$, respectively. 

We treat the nonlinearities in these equations as a perturbation and work in Fourier space in both position and time, $(\mathbf{k},\omega)$. Further, the equations for both $u_i$ and $c$ are diffusive \footnote{The pressure term in Eq.~(\ref{veloeq}) is only used to enforce the incompressibility condition given by Eq.~(\ref{ueq}).} and because $f_i$ and $g$ have identical statistical properties, we focus only on the mathematical structures and scaling solutions.  We refer to the bare transport coefficients $D_0$, and their renormalized values by $D$. Structurally the one-loop contribution to $D$, denoted by $\delta D$, is given by,
\begin{equation}
\begin{split}
&\delta D(\mathbf{k},\omega)\propto\Delta\int_{\mathbf{q},\omega_1}\frac{1}{[-i\omega-i\omega_1+D_0(\mathbf{k}+\mathbf{q})^2][\omega_1^2+D_0^2q^4]} \\
&\propto\Delta\int_{q>(\frac{\omega}{D_0})^{1/2}, \omega_1}\frac{1}{[-i\omega_1+D_0q^2][\omega_1^2+D_0^2q^4]} ,
\end{split}
\label{deltaD}
\end{equation}
where we have assumed that $\Delta\propto\Delta_f\propto\Delta_g$. In going from the first to the second line in Eq.~(\ref{deltaD}) we have replaced the external frequency and wavenumber dependence in the integrand by a frequency cutoff. For $\mathbf{k}=0$, this can be justified in a scaling sense.

Carrying out the frequency and wave number integrals in $d=2$ gives,
\begin{equation}
\delta D(\omega\rightarrow 0)\propto \frac{\Delta}{D_0\omega}
\label{deltaD2}
\end{equation}
Since $\delta D$ diverges as $\omega\rightarrow 0$, we use $\delta D\approx D$ and self-consistently replace $D_0$ in Eq.~(\ref{deltaD2}) by $D$. We conclude that in $d=2$,
\begin{equation}
D(\omega\rightarrow 0)\propto\frac{1}{\omega^{1/2}}.
\end{equation}
This result in turn implies that the velocity autocorrelation function behaves as $C(t\rightarrow\infty)\propto 1/t^{1/2}$ and the mean-squared displacement is $\propto t^{3/2}$. These results are consistent with our numerical work as well as with experiments. In a similar manner,  one obtains $D(\omega\rightarrow 0)\propto\frac{1}{\omega^{1/5}}$, which implies a velocity auto-correlation function $\propto 1/t^{4/5}$  in $d=3$.

\bigskip
\bigskip
\bigskip
\section*{}
 \noindent 
 \textbf{Acknowledgements}
 
\noindent 
 We are grateful to Giorgio Scita et al. for providing us the original experimental data. We thank Davin Jeong for help in producing figures. This work is supported by the National Science Foundation (PHY 17-08128), and the Collie-Welch Chair through the Welch Foundation (F-0019).
\noindent

\noindent 

\noindent 
\noindent 


\noindent 
\textbf{Competing financial interests}

\noindent 
The authors declare no competing financial interests.

\bibliographystyle{naturemag}

\bibliography{cellgrowthinducedvorticity8}

\clearpage
\begin{figure}
{\includegraphics[clip,width=1.1\textwidth]{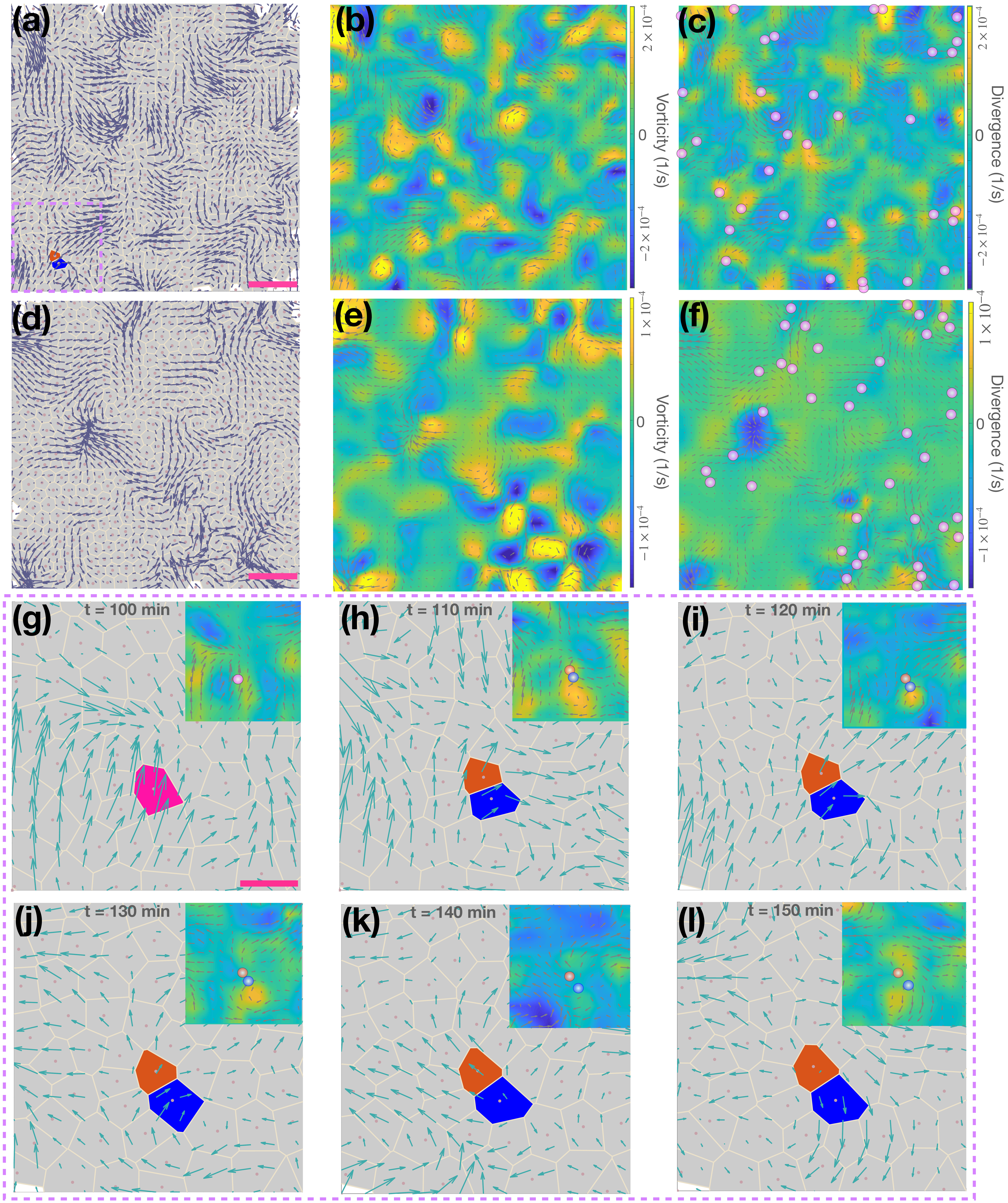}}
\caption{\textbf{Experiments and simulations (an agent-based model) on cell motility in a MDCK monolayer} }
\label{figmodel}
\end{figure}

\clearpage
\begin{figure}
\contcaption{\textbf{Experiments and simulations (an agent-based model) on cell motility in a MDCK monolayer.} \textbf{(a)} A typical velocity vector field in a MDCK cell monolayer, at time $t$ = 120 min,  illustrated by the blue arrows. The positions of the cell nuclei are indicated by the pink dots from which the Voronoi cells are constructed. \textbf{(b)} The vorticity of the cell monolayer for the same time frame as in Fig.~\ref{figmodel}(a). The magnitude and direction of the rotation of the vorticity are coded by the color bar on the right side of the figure. \textbf{(c)} The divergence field of the cell monolayer at $t$ = 120 minute. The magnitude of the divergence is color coded, with outward flux in yellow and inward flux in blue (see the color bar on the right side of the figure). The purple dots show the positions of the new cells, which are born between time $t$ = 95 minute  to $t$ = 120 minute. The velocity vector field is overlaid on the vorticity and divergence fields. Same as \textbf{(a-c)}, the velocity vector \textbf{(d)}, vorticity \textbf{(e)}, and divergence \textbf{(f)} fields calculated from simulations.  The pink bar in  \textbf{(a)} and  \textbf{(d)} corresponds to 50 $\mu m$. The same time interval (25 minutes) as in  \textbf{(c)} is used in \textbf{(f)} for the positions of the new cells.   \textbf{(g-l)} A zoom in of the dashed rectangle region at the lower left corner of Fig.~\ref{figmodel}\textbf{(a)} for different times, which shows the formation of a vortex accompanied by a single cell division event.  Cell division is observed in Fig.~\ref{figmodel}\textbf{(h)} (see the orange and blue cells originated from the pink cell division in Fig.~\ref{figmodel}\textbf{(g)}), and  a vortex  is formed that accompanied by such an event (see the velocity field close to the blue cell). To illustrate the formation of the vortex vividly, the vorticity field, as in Fig.~\ref{figmodel}\textbf{(b)}, is also shown in the inset of each figure at the upper right corner (see the yellow regions in Fig.~\ref{figmodel}\textbf{(h-j)}, especially). The positions of the mother and daughter cells are indicated by dots in different colors. The pink bar in  \textbf{(g)} corresponds to 20 $\mu m$.  }  
\end{figure}
\clearpage

\clearpage
\begin{figure}
{\includegraphics[clip,width=1.05\textwidth]{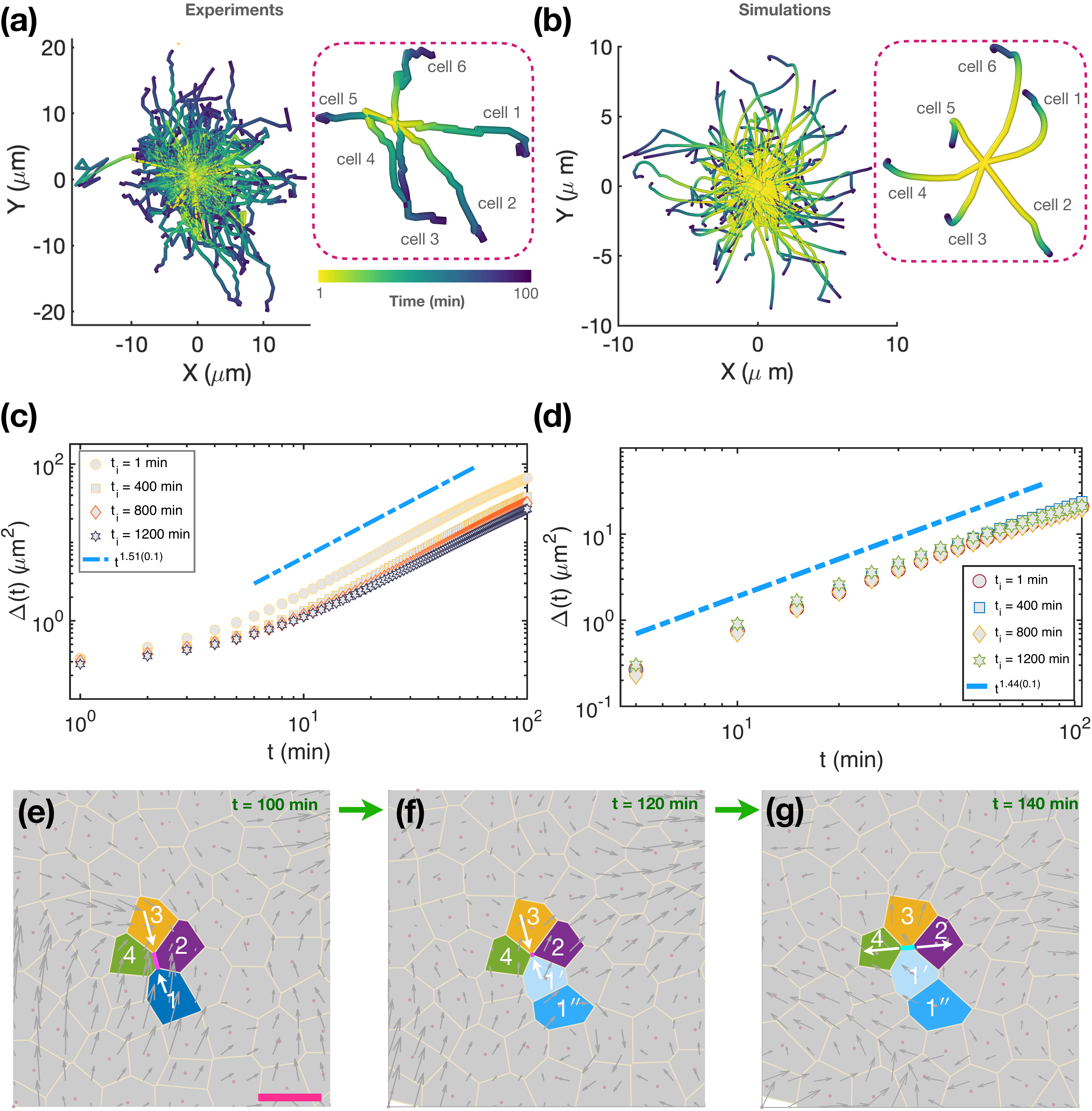}}
\caption{\label{msddis}\textbf{Anomalous diffusion and topological changes.}}
\end{figure}


\clearpage
\begin{figure}
\contcaption{ \textbf{Anomalous diffusion and topological changes.}   
\textbf{(a)} The trajectories of the first 150 newly generated cells for the initial 100 minutes after birth. The initial position of the new cells are aligned at the origin (0,0) for illustration purpose only. Inset on the right:  Six randomly chosen representative trajectories from \textbf{(a)}. \textbf{(b)} Same as \textbf{(a)}, except they are from simulations.  The trajectories in \textbf{(a)-(b)} are color coded by time (scale on the bottom of the inset in \textbf{(a)}).  \textbf{(c)}  Mean-square displacement ($\Delta (t)$) of cells calculated over a time window of 200 minutes, starting from different initial time points $t_i$ over the course of the experiment\cite{giavazzi2018tracking}. \textbf{(d)} Same as \textbf{(c)}, except the results are from simulations. The same time window of 200 minutes is used for each $t_i$.  The slope (1.44) of the dashed-dotted line from simulations is similar as shown in \textbf{(c)} for experiments (1.51), both showing superdiffusive behavior.   \textbf{(e)-(g)} T1 transition induced by cell divisions. Cell $1$ in \textbf{(e)} divided into two cells (cell $1'$, and cell  $1''$) in \textbf{(f)}. The motion of newly generated cells leads to T1 transition, which separates the two neighboring cells (cell 2 and cell 4) away from each other in \textbf{(g)} (see also Fig.~S7 in the SI for more details). The pink bar in  \textbf{(e)} corresponds to 20 $\mu m$. Same as Figs.~\ref{figmodel}\textbf{g-l}, a vortex around cells $1'$ and $1''$ is found in \textbf{(f)} (see the velocity vector field in grey arrows).}  
\end{figure}
\clearpage


\clearpage
\begin{figure}
{\includegraphics[clip,width=1.05\textwidth]{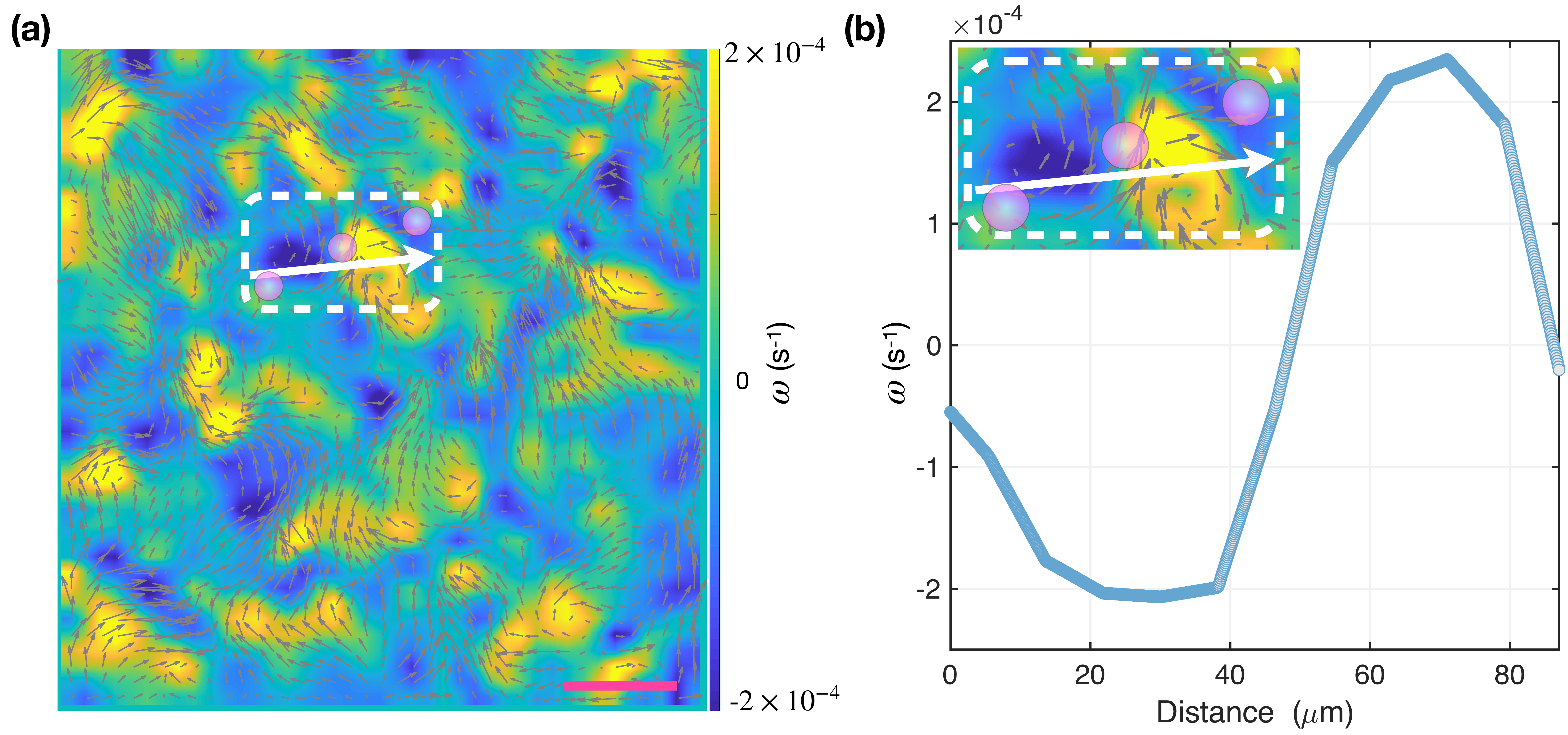}}
\caption{\textbf{Clockwise and anticlockwise vorticity ($\omega$).} \textbf{(a)} A pair of vortices with clockwise and anticlockwise rotations at time $t$ = 80 minute using the vorticity field of a cell monolayer (see the region indicated by the dashed rectangle).  Three purple circles mark three cells, which were born between $t$ = $70-80$ minutes. The velocity vector field is overlaid on the vorticity fields.  The pink bar corresponds to 50 $\mu m$. Scale on the right gives $\omega$ in unit of $s^{-1}$. \textbf{(b)} Vorticity as a function of the distance from the left side of the dashed box along the white arrow in \textbf{(a)}. A zoom in of the dashed rectangle in \textbf{(a)} is  shown in the inset.}
\label{clock}
\end{figure}

\clearpage
\floatsetup[figure]{style=plain,subcapbesideposition=top}
\begin{figure}
{\includegraphics[clip,width=1.05\textwidth]{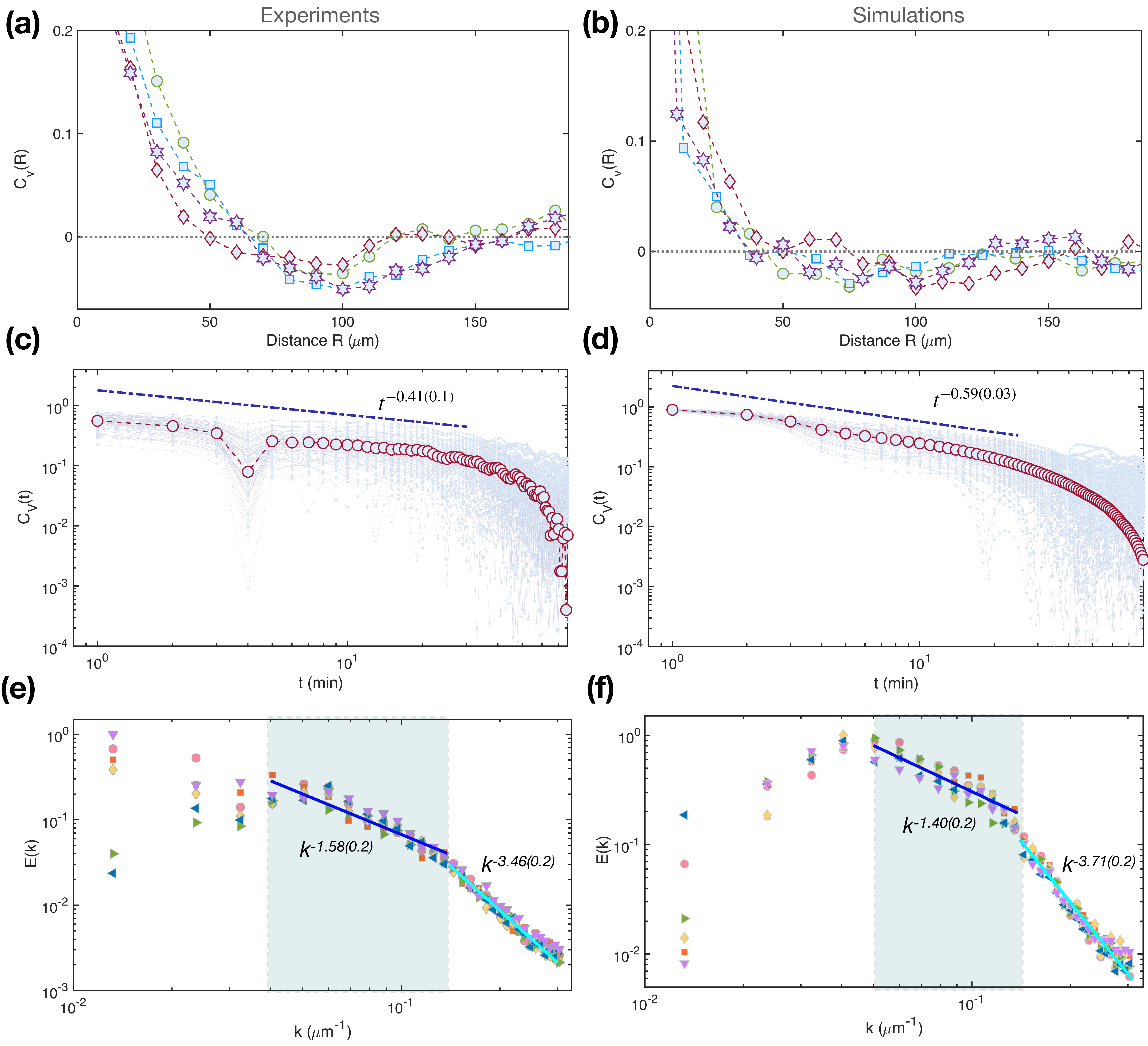}}
\caption{\label{cvt}\textbf{Velocity autocorrelation and energy spectra.}}
\end{figure}

\clearpage
\begin{figure}
\contcaption{ \textbf{Velocity autocorrelation and energy spectra.} \textbf{(a)} The velocity spatial correlation ($C_v(R)$) at different times: $t_i$ = 100 (cirlces), 600 (squares), 1200 (diamonds), and 1400 (hexagrams) minutes. Each curve is averaged over 10 time frames. A negative minimum value at $R_v \approx 100~\mu m$ ($\approx 10$ times the cell size) is present in all the curves.   \textbf{(b)} Same as  \textbf{(a)}, except it is obtained from simulations.  We took $\delta t = 10$ minutes in \textbf{(a-b)} and the dotted lines show the zero value. \textbf{(c)} The velocity autocorrelation ($C_{v}(t)$) as a function of time in a double logarithmic scale calculated using $\delta t =4$ minutes (see Fig.~S12 in the SI for different values of $\delta t$).  The data from a single cell analysis (135 cells in total which are alive through the whole experiment) is shown in grey dots and the red circles show the mean value. The slope of the dash-dotted line is -0.41$\pm$0.1.   \textbf{(d)} Same as \textbf{(c)}, except the result is obtained from simulations. The slope of the dashed-dotted line is -0.59$\pm$0.03.   \textbf{(e)} The energy spectra $E(k)$ calculated using the trajectories generated in imaging experiments at different times: $t$ = 100 (cirlces), 110 (squares), 120 (diamonds), 130 (right-pointing triangle), 140 (left-pointing triangle) and 150 (down-pointing triangle) minutes.  \textbf{(f)} Calculated $E(k)$ from simulations. The values of the slope of the solid lines are shown in the figure.  In the grey regime $E(k) \sim k^{-1.58}$ from experiments. This value (1.58) is hauntingly close to 5/3 expected for inertial turbulence. The two exponent values obtained from simulations (1.4 and 3.7) are close to the values obtained from analysis of the data from experiments (1.6 and 3.5). The unit for the wave vector ($k$) is $\mu$m$^{-1}$, and $E(k)$ ($\mu$m$^3$/s$^2$) is scaled by the maximum value. }
\end{figure}

\noindent 
\end{document}


\title{{Topological transitions, turbulent-like motion and long-time-tails} driven by cell division in biological tissues - Supplementary Information}


\author{Xin Li}
\affiliation{Department of Chemistry, University of Texas, Austin, TX 78712, USA.}
\author{Sumit Sinha}
\affiliation{Department of Physics, University of Texas, Austin, TX 78712, USA.}
\author{ T. R. Kirkpatrick}
\affiliation{lnstitute of Physical Science and Technology, University of Maryland, College Park, Maryland 20742}
\author{D. Thirumalai}
\email{dave.thirumalai@gmail.com}
\affiliation{Department of Chemistry and Department of Physics, University of Texas, Austin, TX 78712, USA.}
\date{\today}

\begin{abstract}

\end{abstract}

\pacs{}

\maketitle

\section*{}
In the Supplementary Information (SI), we provide a summary of experimental results, several additional findings, description of the simulations using a particle-based model, and related discussions that are pertinent to the results present in the main text.

\textbf{Microscopy Experiments:} 
Because the present study uses results from the imaging experiments\cite{Giavazzi2018}, we describe them in some detail. Two types of microscopes,  phase-contrast and wide-field fluorescence microscopy are used in the MDCK cell experiments. A confluent monolayer, with cells distributed evenly over the whole space, is imaged using phase contrast microscopy, allowing the capture of the cell membrane (see Fig.~\ref{exfigure}a and movie 2 in the supplementary materials of the original article\cite{Giavazzi2018}).  In addition, Enhanced Green Fluorescent Protein (EGFP)-tagged H2B histones were constitutively expressed by retroviral infection using standard protocols\cite{Giavazzi2018}. The images of the nuclei of all the cells, at each time frame, were captured by the wide-field fluorescence microscopy (see Fig.~\ref{exfigure}b and also the movie 1 in the supplementary materials in the experiments\cite{Giavazzi2018}).  Cell division occurred frequently (see Fig.~\ref{exfigure}c). The left panel of Fig.~\ref{exfigure}c shows an example of a cell division event using the phase-contrast microscopy, while the right panel illustrates the same event using the wide-field fluorescence microscopy. The protocol in the experiments \cite{Giavazzi2018} was designed to avoid any directional bias for cell motility patterns in order to reduce the inhomogeneity and anisotropy of the cell density and shape in the monolayer. Using MosaicSuite plugin of the ImageJ\cite{Rueden2017,Schindelin12} for single particle tracking over the fluorescence images of the cell nuclei (one-minute resolution over 24 hours), the positions of all the cells were  recorded in consecutive frames. Finally, the trajectories of all the cells in one field-of-view (FOV) could be reconstructed (see Fig.~\ref{exfigure}d). The data from the same FOV, as in Fig.~\ref{exfigure}d, was used for the analyses presented in the main text and the SI.


\textbf{Temporal evolution of a cell monolayer:} First, we explore some general features of the temporal evolution of the MDCK monolayer using the experimentally generated trajectories\cite{Giavazzi2018}.  Although the MDCK monolayer is confluent initially, it does not stay in a stationary state but evolves with time,  exhibiting rich dynamics driven by cell division and apoptosis. 


{\it Cell divisions and apoptosis:}
The number of cell divisions (see Fig.~\ref{exfigure}c) between two frames (one minute interval) is quantified in Fig.~\ref{newcell}a (see the green circles). Cell apoptosis/extrusion events are also shown in the same figure by yellow stars. These two numbers fluctuate between 0 and 12. Therefore, the confluent cell monolayer is  dynamic that  undergoes continuous cell division and apoptosis, which generate active forces leading to the collective cell motions\cite{Sinha20,Doostmohammadi2015,Rossen2014}. The mean value for the number of new cells (5.14) is a slightly higher than the dead cells (5.05). Hence, the number of cells in the FOV increases with time, as shown in Fig.~\ref{newcell}b.  A power-law relation describes the cell number as a function of time that is shown by the green solid line and the expression is given in the figure caption.

{\it Cell size distributions:}
Due to cell growth, division and apoptosis, the cell size is not uniformly distributed. In order to calculate the distribution of cell sizes, we used Voronoi tessellation of the positions of all the cell nuclei in the FOV (Fig.~\ref{exfigure}d). The Voronoi cells at two different time frames are shown in Figs.~\ref{voronoicell}a-b, and the area of each cell is color coded as indicated in the bar on the right side of the figures.  There is a large variation (from 100 to 400 $\mu m^2$) in the cell area. The cell area distributions, $P(A)$, at  early (sampled over 20 images from $t$ = 180 to 199 minutes) and later (1350-1369 minutes) times are illustrated in Fig.~\ref{voronoicell}c (without considering the boundary Voronoi cells). A Gaussian distribution (see the solid lines and the functional forms shown in Fig.~\ref{voronoicell}c) describes the $P(A)$ reasonably well. The mean value of the cell area decreases from 231  $\mu m^2$ at early times to 187 $\mu m^2$ at later times during a 24-hour period in the experiments. This is an indication of cell jamming \cite{Atia2018,Bi2016}.

{\it Cell density and velocity:}
As the number of cells increases (see Fig.~\ref{newcell}b) in the same FOV, so does the cell density (see Fig.~\ref{velocityvstime}a, defined as the ratio of the cell number and the area in the FOV). A similar power-law relation is found for the cell density and the cell number (see the solid lines and functional forms shown in Fig.~\ref{newcell}b and Fig.~\ref{velocityvstime}a).  The mean velocity, defined in the main text above Eq.~(1) with $\delta t = 10$ min) of cells, decreases with time (see the green line in Fig.~\ref{velocityvstime}b). Therefore, the overall motility of cells is inhibited as the cell density increases, even though the frequencies of active cell division and apoptosis  are nearly constant (see Fig.~\ref{newcell}a).

{\it Distribution of cell coordination number: }
The number of neighbors (coordination number, $Z$) per cell in MDCK monolayer varies from  cell to cell (see Figs.~\ref{voronoicell}a-b) because of the generation of active forces. From the Voronoi tessellation (Fig.~\ref{voronoicell}), we calculated the cell coordination number distribution, $P(Z)$, for the MDCK cell monolayer (see the histograms in blue in Fig.~\ref{numofneighboringcells}). Each histogram is calculated using ten time frames from the experiments\cite{Giavazzi2018}. The value of $Z$ varies from 4 to 9, and its distribution is well described by a Gaussian distribution with a mean value around 6 (see the solid lines), which remains robust during the 24 hours experiment (see Figs.~\ref{numofneighboringcells}a-b). The distribution, $P(Z)$, obtained from our simulations agrees well with the experimental results (see the histograms and also the solid lines in Fig.~\ref{numofneighboringcells}).  Therefore, a predominantly hexagonal pattern can be driven purely by cell division and apoptosis, which leads to an emergent phenomena of geometric order,  similar to that found in proliferating metazoan epithelia\cite{gibson2006emergence}.

\textbf{Displacement distribution of cells: }
We calculated the displacement distributions ($P(\Delta x)$, $P(\Delta y)$) of cells  using the trajectories from the  MDCK cell experiments and our simulations (see Fig.~\ref{displacemsd}). Dramatic deviations  from the Gaussian distribution (see the green solid lines) are found in both experiments and simulations.

\textbf{T2 transition induced by cell apoptosis: }
The T1 transition induced by cell divisions is illustrated in Figs.~2e-g in the main text. We also found evidence for another type of topological change, the T2 transition, caused by cell apoptosis/extrusion from the confluent monolayer (see Fig.~\ref{T2transitionfinal}). Through Voronoi tessellation, using the cell nuclei position data, we found that cell $0$ (the one shown in red) is extruded from the cell monolayer at a later time (see Fig.~\ref{T2transitionfinal}b), leading to the formation of new neighborhood constituting cells $1-6$. For completeness, we should point out that T1 and T2 transitions are well known in the dynamics of foams  \cite{Weaire1984} where connections to cells was also explained.

\textbf{Velocity autocorrelation at different $\delta t$: }
We showed that  the velocity autocorrelation function exhibits long time tail (LTT) behavior  in the main text (see Fig.~4c). To ascertain whether LTT depends on the time interval, $\delta t$,  we calculated velocity autocorrelation of MDCK cells \cite{Giavazzi2018} at several values of $\delta t$ (2, 4, and 10 minutes). The result is robust (see the different symbols and the dash-dotted lines in Fig.~\ref{cvtm}). 

\textbf{Energy spectrum:}
The kinetic energy $E$ is given by,
\begin{equation}
E = \frac{1}{2} \int |\textbf{v}(\textbf{r})|^2 d\textbf{r} = \int E(\textbf{k}) d\textbf{k},
\end{equation} 
where $\textbf{v}(\textbf{r})$ is the cell velocity at position $\textbf{r}$, and $k = |\textbf{k}|$ is the wave number. To calculate the energy spectrum $E(k)$ at a time point, we first use the digital particle image velocimetry\cite{Thielicke2014} to get the velocity field $\textbf{v}(\textbf{r})$ (using two images with $\delta t = 10$ minutes) at regular grid points. Then, we used the fast Fourier transform (FFT) to convert the velocity field $\textbf{v}(\textbf{r})$ to $\tilde{\textbf{v}}(\textbf{k})$. Finally, the energy spectrum $E(k)$ is calculated using,
\begin{equation}
E(k) = \frac{kA_0}{4\pi}\sum \tilde{\textbf{v}}^*(\textbf{k}) \cdot \tilde{\textbf{v}}(\textbf{k}) ,
\end{equation} 
where $\tilde{\textbf{v}}^*(\textbf{k}) $ is the complex conjugate of $\tilde{\textbf{v}}(\textbf{k})$ and $A_0$ is the area of the FOV. Two examples for $E(k)$ are illustrated here: one is the Fig.~4e in the main text and the other is shown in Fig.~\ref{EK500}. In each figure, we calculated the $E(k)$ at six different time frames. The dependence of $E(k)$ on $k$ shows that there are two scaling regimes (see the solid lines and also the values of the slope).  The two solid lines give the best power-law fit over the data (all six time frames) for different regimes of the wavevector $k$.

\textbf{Agent-based simulations:}
We use a two dimensional (2D) version of the three dimensional (3D) agent-based model\cite{Malmi2018,Sinha2020}. In 2D, the cells are discs. We just give a sketch of the model because additional details can be found elsewhere\cite{Malmi2018}. We use a deformable disc to represent each cell. The cell area can increase at a constant rate ($r_{A}$) as long as the pressure, $p$, it experiences due to contact with the neighboring cells is less than a preassigned critical value $p_{c}$\cite{Malmi2018}.  If $p$ exceeds $p_c$, the cell enters a dormant state, stops growing unless the normal pressure is below $p_{c}$, which could occur as the system evolves. The use of $p_{c}$ accounts for  cell proliferation inhibition by a mechanical feedback mechanism\cite{Shraiman2005,Montel2011,Quail2013,Jain2014,Delarue2016}.  After reaching the fixed mitotic radius $R_{m}$, with  $p<p_{c}$, a cell divides into two  with identical radius, $R_d = R_{m}2^{-1/2}$ ($\pi R_m^2 = 2 \pi R_d^2$). The new cells are placed randomly in space (the mother cell occupied before division) at a center-to-center distance $d = 2R_{m}(1-2^{-1/2})$.  The cell area grows at a rate $r_{A} = \pi R_m^2/(2\tau_{min})$ ($\tau_{min}$ is the cell cycle time). We update the cell radius ($R$) from a Gaussian distribution with the mean value $\dot R = r_{A}/(2\pi R)$.

The elastic force $F_{ij}^{el}$ between two cells of radii $R_{i}$ and $R_{j}$ are modeled by Hertzian contact mechanics~\cite{schaller2005multicellular, pathmanathan2009computational, drasdo2005single},
\begin{equation}
\label{rep}
F_{ij}^{el} = \frac{h_{ij}^{3/2}(t)}{\frac{3}{4}(\frac{1-\nu_{i}^2}{E_i} + \frac{1-\nu_{j}^2}{E_j})\sqrt{\frac{1}{R_{i}(t)}+ \frac{1}{R_{j}(t)}}},
\end{equation}
where $E_{i}$ and $\nu_{i}$ are the elastic modulus and
Poisson ratio of cell $i$, respectively. The overlap $h_{ij}$ between the two cells is defined as $\mathrm{max}[0, R_i + R_j - |\vec{r}_i - \vec{r}_j|]$ with $|\vec{r}_i - \vec{r}_j| \equiv r_{ij}$,  being the center-to-center distance of the cells.  In the simulations, we use $E_i$ and $\nu_{i}$ to  be independent of $i$. Cell-cell adhesive interaction, $F_{ij}^{ad}$, is mediated by receptor and ligand proteins on the cell membrane which is calculated using\cite{schaller2005multicellular},
\begin{equation}
\label{ad}
F_{ij}^{ad} = L_{ij}f^{ad}\frac{1}{2}(c_{i}^{rec}c_{j}^{lig} + c_{j}^{rec}c_{i}^{lig}),
\end{equation}
where the $c_{i}^{rec}$ ($c_{i}^{lig}$) is the receptor (ligand) concentration, which is considered to be constant here,  and  $f^{ad}$ is a coupling constant used to rescale the adhesion force.  
The length of the contact line, $L_{ij}$ between two cells, is given by $L_{ij} = \sqrt{|4r_{ij}^2 R_{i}^2-(r_{ij}^2-R_{j}^2+R_{i}^2)^2 |}/{r_{ij}}$. The total force on the $i^{th}$ cell is  the sum over its nearest neighbors ($NN(i)$), 
\begin{equation}
\vec{F}_{i} = \Sigma_{j \epsilon NN(i)}(F_{ij}^{el}-F_{ij}^{ad})\vec{n}_{ij}. 
\end{equation}
where $\vec{n}_{ij}$ is  the unit vector pointing from the center of cell $j$ to the center of cell $i$.

The spatial dynamics of the cell is calculated by integrating the equation of motion for a cell of mass $m_{i}$,
\begin{equation}
\label{eom}
m_{i}\ddot{\vec{r}}_{i} = \vec{F}_{i}(t) -  \gamma_{i} \dot{\vec{r}}_{i}(t),          
\end{equation}
where  $\gamma_{i}$ is the modified mobility coefficient of cells with  $\gamma_{i}= \gamma_{i}^{visc} + \gamma_{i}^{ad}$.
The first term comes from the contribution of cell to extracellular matrix (ECM) friction ($\eta r_{i}$, with $\eta$ the mobility constant)  and the second term describes the  cell-to-cell friction with,
\begin{eqnarray}
\gamma_{i}^{ad} =&& \gamma^{max}\Sigma_{j \epsilon NN(i)} [L_{ij}\frac{1}{2}(1+\frac{\vec{F}_{i} \cdot \vec{n}_{ij}}{|\vec{F}_{i}|})\times \\ \nonumber &&\frac{1}{2}(c_{i}^{rec}c_{j}^{lig} + c_{j}^{rec}c_{i}^{lig})] \, .
\end{eqnarray}
Because Reynolds number for cells in a tissue is low~\cite{dallon2004cellular}, the inertial term in Eq.~(\ref{eom}) can be neglected~\cite{schaller2005multicellular}. Therefore, the equation of motion becomes, 
\begin{equation}
\label{eqforce}
\dot{\vec{r}}_{i} = \frac{\vec{F}_{i}}{\gamma_i}.
\end{equation}
We calculated the pressure $p$ as in previous studies (Eq.~(8) in Ref.\cite{Malmi2018}) except the contact line length, $L_{ij}$, is used instead of the contact area, $A_{ij}$, between two cells. We employed periodic boundary conditions instead of the free boundary for a growing tumor spheroid\cite{Malmi2018,Sinha2020}. 

We start from 300 cells that are randomly distributed in a box with the size $340\mu m \times 340\mu m$, similar to the one in the FOV in the experiments\cite{Giavazzi2018} (see Fig.~1 in the main text). The parameters in the agent-based model used in the simulations are given in Table~\ref{tableparameter}. Note that the ratio of the rate of apoptosis to cell division ($0.54$) is comparable to our estimation using the experimental data.

\clearpage
\newpage
\begin{table}[ht!]
\caption{The parameters used in the simulation.}
\begin{tabular}{lcr}
\toprule[1.5pt]
 \bf{Parameters} & \bf{Values} & \bf{References} \\
 \hline
 Time step ($\Delta t$)&  10 $\mathrm{s}$  &  ~\cite{Malmi2018} \\
Critical Radius for Division ($R_{m}$) &  11 $\mathrm{\mu m}$ & This paper\\
 Mobility constant ($\eta$) & 0.1 $\mathrm{kg/ (\mu m~s)}$   & This paper  \\
 Benchmark Cell Cycle Time ($\tau_{min}$)  & 54000 $\mathrm{s}$  & ~\cite{Freyer1986,Casciari1992,Landry1981}\\
 Adhesive Coefficient ($f^{ad})$&  $10^{-4} \mathrm{\mu N/\mu m}$  & ~\cite{schaller2005multicellular} \\
Mean Cell Elastic Modulus ($E_{i}) $ & $10^{-3} \mathrm{MPa}$  & ~\cite{Galle2005}    \\
Mean Cell Poisson Ratio ($\nu_{i}$) & 0.5 & ~\cite{schaller2005multicellular}  \\
 Death Rate ($b$) & $10^{-5} \mathrm{s^{-1}}$ & This paper \\
Mean Receptor Concentration ($c^{rec}$) & 1.0 (Normalized) & ~\cite{schaller2005multicellular} \\
Mean Ligand Concentration ($c^{lig}$) & 1.0 (Normalized) & ~\cite{schaller2005multicellular}  \\
Adhesive Friction $\gamma^{max}$ &  $10^{-4} \mathrm{kg/ (\mu m~s)}$  &  ~\cite{Malmi2018}\\
Threshold Pressue ($p_c$) & $5\times10^{-3} \mathrm{\mu N/\mu m}$  & This paper    \\
\bottomrule[1.5pt] 
\end{tabular}
\label{tableparameter}
\end{table}

\clearpage
\begin{figure}
{\includegraphics[clip,width=0.85\textwidth]{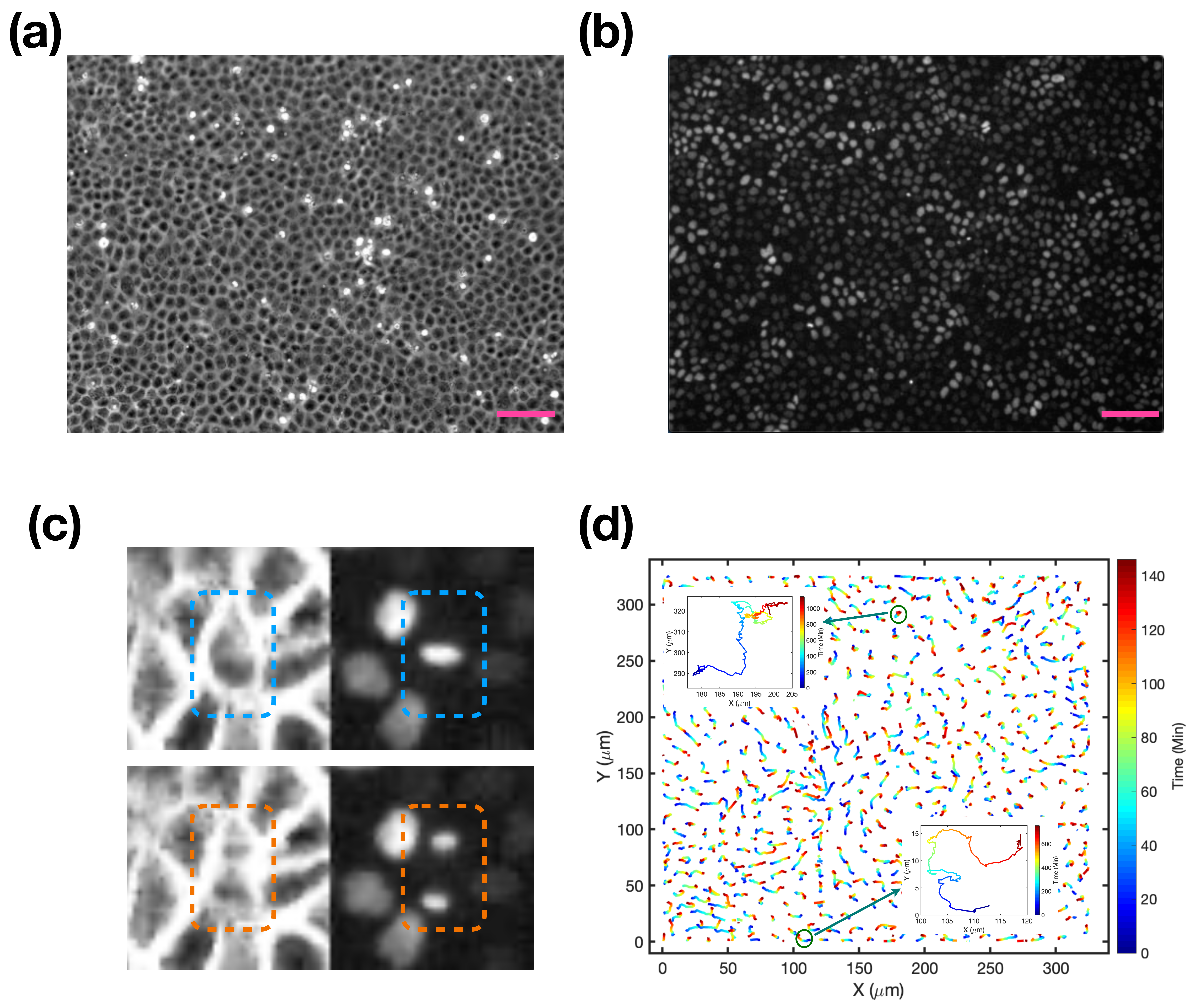}}
\caption{\textbf{MDCK monolayer.} (a) A representative image of a confluent MDCK monolayer (t = 0 min) using the phase-contrast microscopy (taken from experiments by Fabio et al.\cite{Giavazzi2018}), which highlights the cell membrane. (b) Same as (a) except the image is taken using the wide-field fluorescence microscopy\cite{Giavazzi2018}, where the cell nuclei are highlighted. The size of the images in (a)-(b) is (867 $\times$ 660) $\mu m^2$, and the pink bars correspond to 100 $\mu m$. (c) A cell division event\cite{Giavazzi2018}. Upper (lower) panel: before (after)  a cell division. The left panel are images from the phase-contrast microscopy. Right panel: images from the wide-field fluorescence microscopy for EGFP-H2B expressing cells.  The size of the image is (60 $\times$ 60) $\mu m^2$. (d) The cell trajectories in one field of view (FOV) are shown over the first 150 minutes. The trajectory is color coded by time (see the color bar on the right of the figure). The insets display two representative cell trajectories over 24 hours. Heterogeneity in the dynamics and swirl patterns are evident.}
\label{exfigure}
\end{figure}

\clearpage
\floatsetup[figure]{style=plain,subcapbesideposition=top}
\begin{figure}
	\sidesubfloat[]{\includegraphics[width=1.0\textwidth] {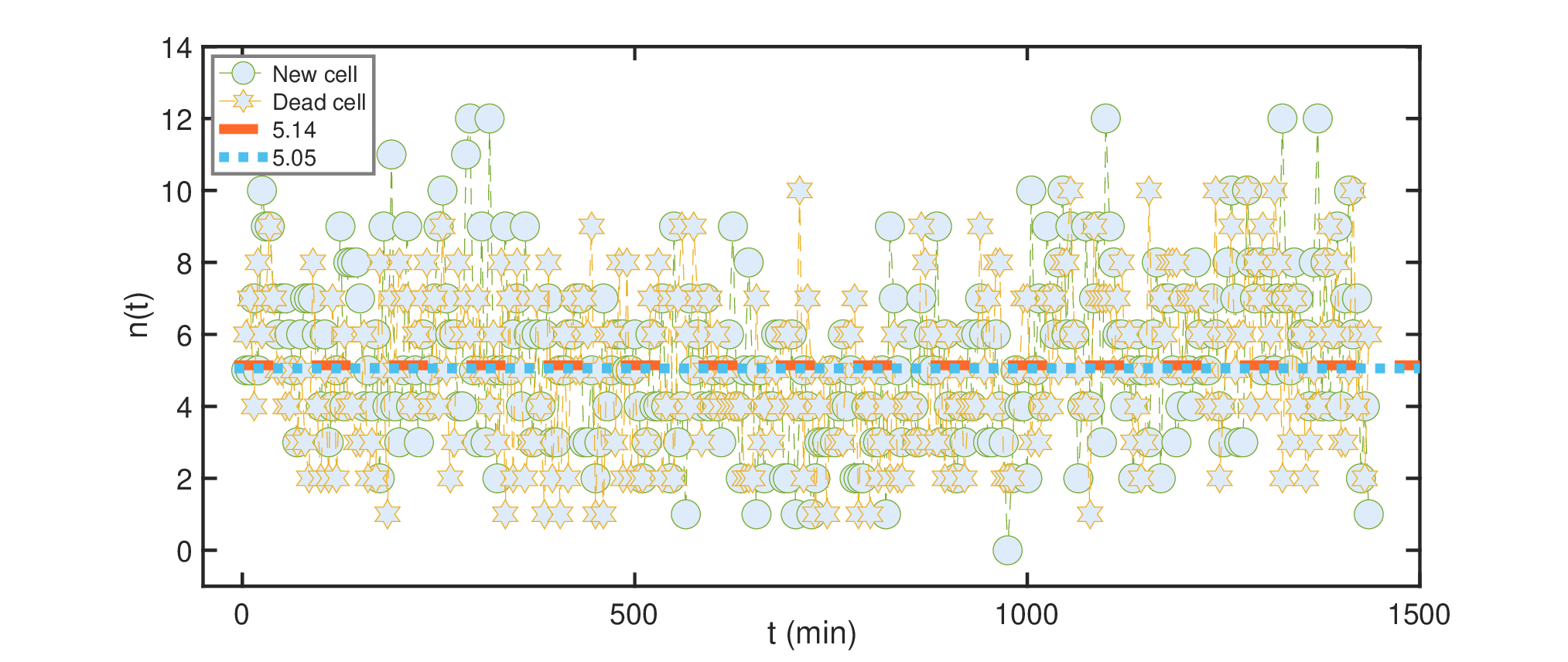} \label{newcell-deadcellnumber} }
						\par\bigskip
		\sidesubfloat[]{\includegraphics[width=1.0\textwidth] {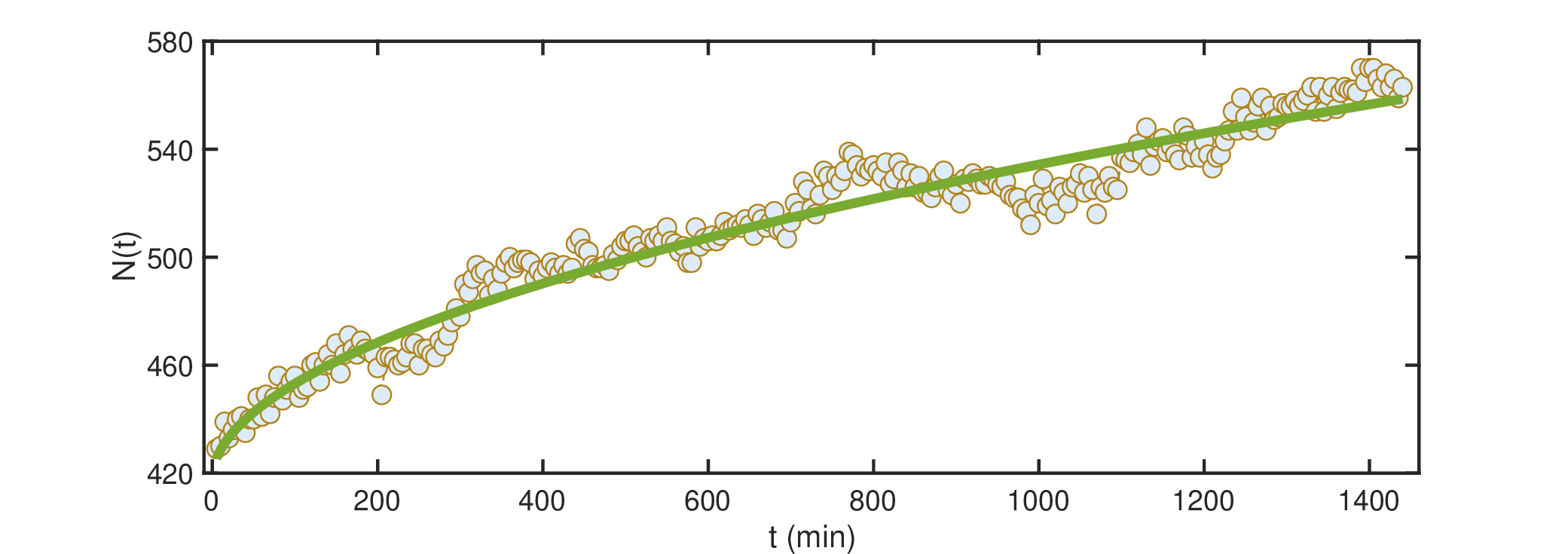} \label{cellnumfinal} }
\caption{\textbf{Cell number dynamics.} (a) The number of new and dead or extruded cells ($n(t)$) in one FOV (see Fig.~\ref{exfigure}d) over the duration of the experiment. The red dashed and blue dotted lines are the mean values for the new and dead/extruded cells, respectively. (b) The temporal evolution of the cell number ($N(t)$ in the same FOV as (a). The green curve is a power law  ($N(t) = 418+ 3.2t^{0.5}$) fit of the experimental data in circles. }
\label{newcell}
\end{figure}


\clearpage
\floatsetup[figure]{style=plain,subcapbesideposition=top}
\begin{figure}
	\sidesubfloat[]{\includegraphics[width=0.50\textwidth] {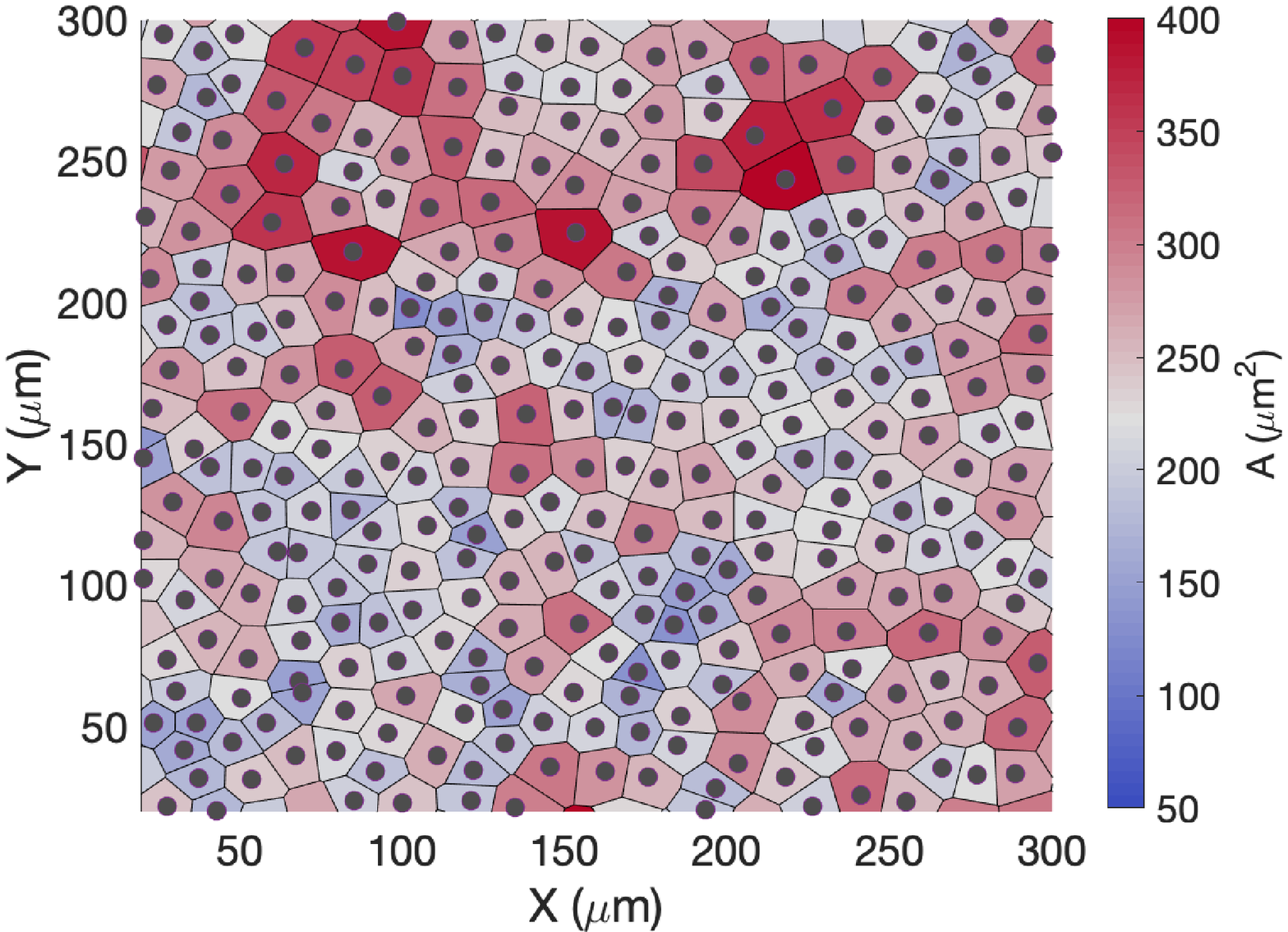} \label{voronoi_180min_nucleus} }
	\sidesubfloat[]{\includegraphics[width=0.50\textwidth] {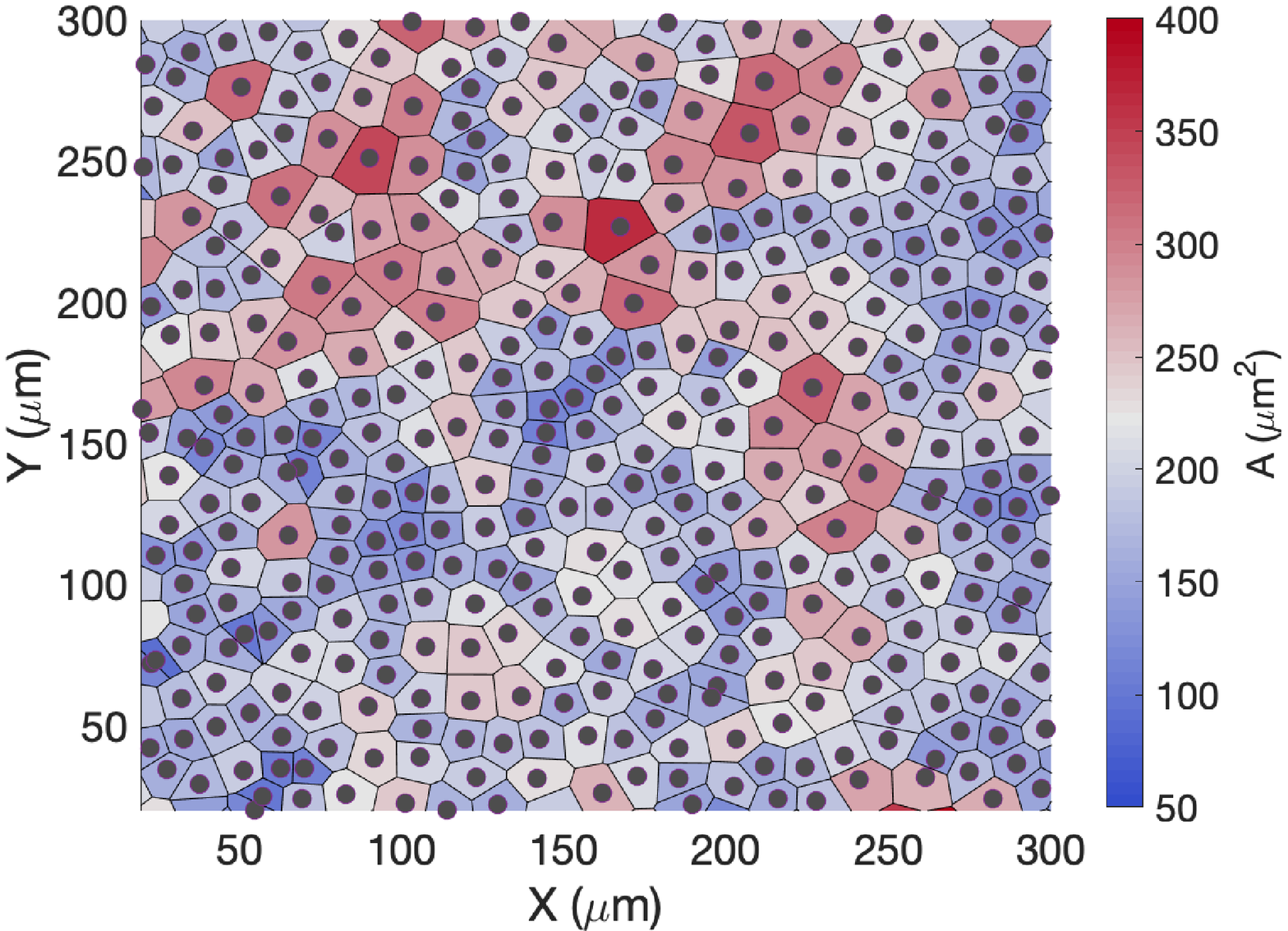} \label{voronoi_1350min_nucleus} }
		\par\bigskip
		\sidesubfloat[]{\includegraphics[width=0.950\textwidth] {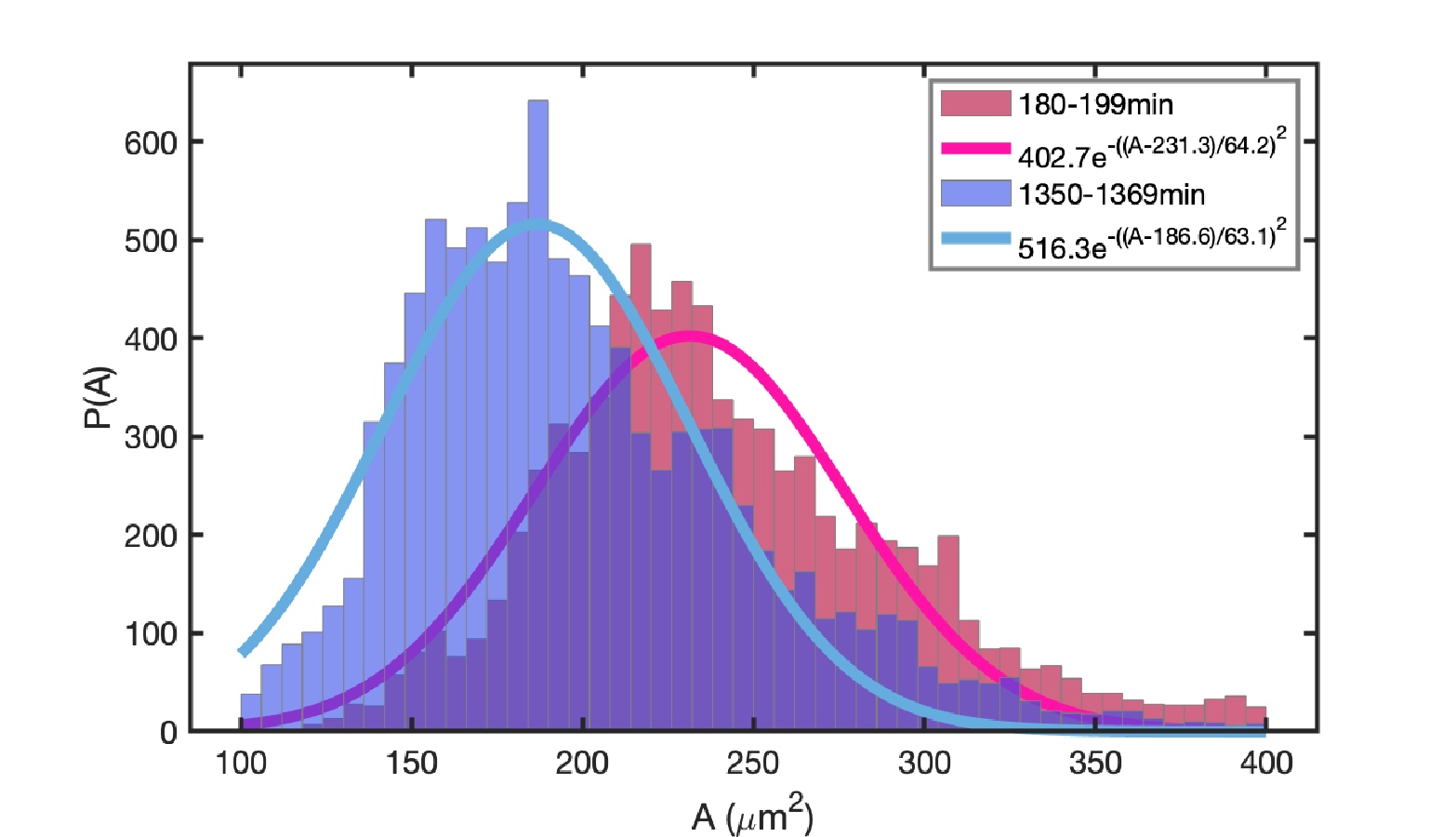} \label{areadis-180-199-1350-1369} }
\caption{\textbf{Cell size distributions of epithelial monolayers.} (a) A representative image of a confluent cell monolayer at an early time ($t$ = 180 minute). (b) Same as (a) except  $t$ = 1350 minutes in the experiments. In (a) and (b), the size of each Voronoi cell is color coded according to the area of each cell (see the bar on the right of the figure). (c) The distribution $P(A)$ of the cell area ($A$) between $t=180$ to $t=199$ minutes (blue), and  between $t=1350$ and $t=1369$ minutes (pink). The solid lines show the best fit to the data. The functional forms are in the inset. }
\label{voronoicell}
\end{figure}

\clearpage
\floatsetup[figure]{style=plain,subcapbesideposition=top}
\begin{figure}
		\sidesubfloat[b]{\includegraphics[width=1.0\textwidth] {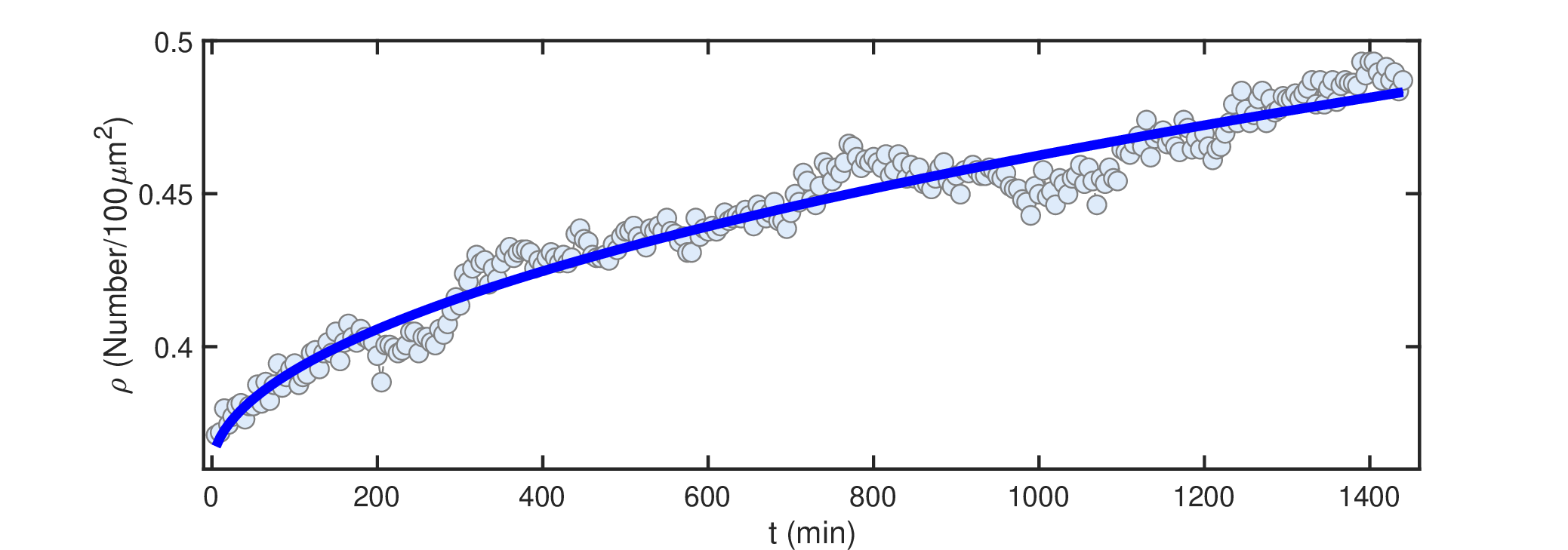} \label{celldensityfinalwithoutaddingdata} }
		\par\bigskip
		\sidesubfloat[b]{\includegraphics[width=1.0\textwidth] {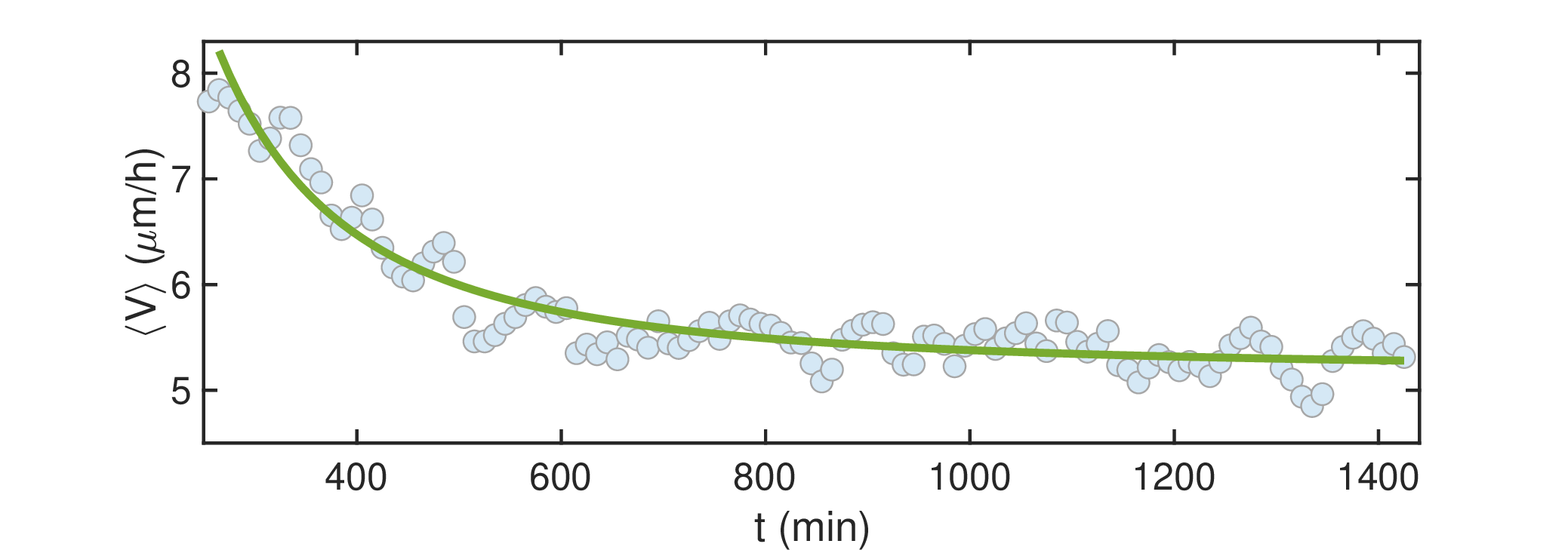} \label{velocityvstime} }
\caption{\textbf{Cell number density ($\rho$) and mean velocity ($\langle V\rangle$).} (a) The evolution of $\rho$ in unit of number per 100 $\mu m^2$. The blue line is the power law fit ($\rho = 0.36+0.003t^{0.5}$) of the experimental data, shown in circles. (b) The mean velocity of cells as a function of time. Each circle represents the mean velocity of all the cells in ten time frames. The green line, which is the power law fit ($\langle V\rangle =5.2+3.3\times10^5 t^{-2.05}$) of  the experimental data, shows a decrease as the cell density increases, which is an indication of cell jamming.  }
\label{velocityvstime}
\end{figure}


\clearpage
\floatsetup[figure]{style=plain,subcapbesideposition=top}
\begin{figure}
		\sidesubfloat[b]{\includegraphics[width=1.0\textwidth] {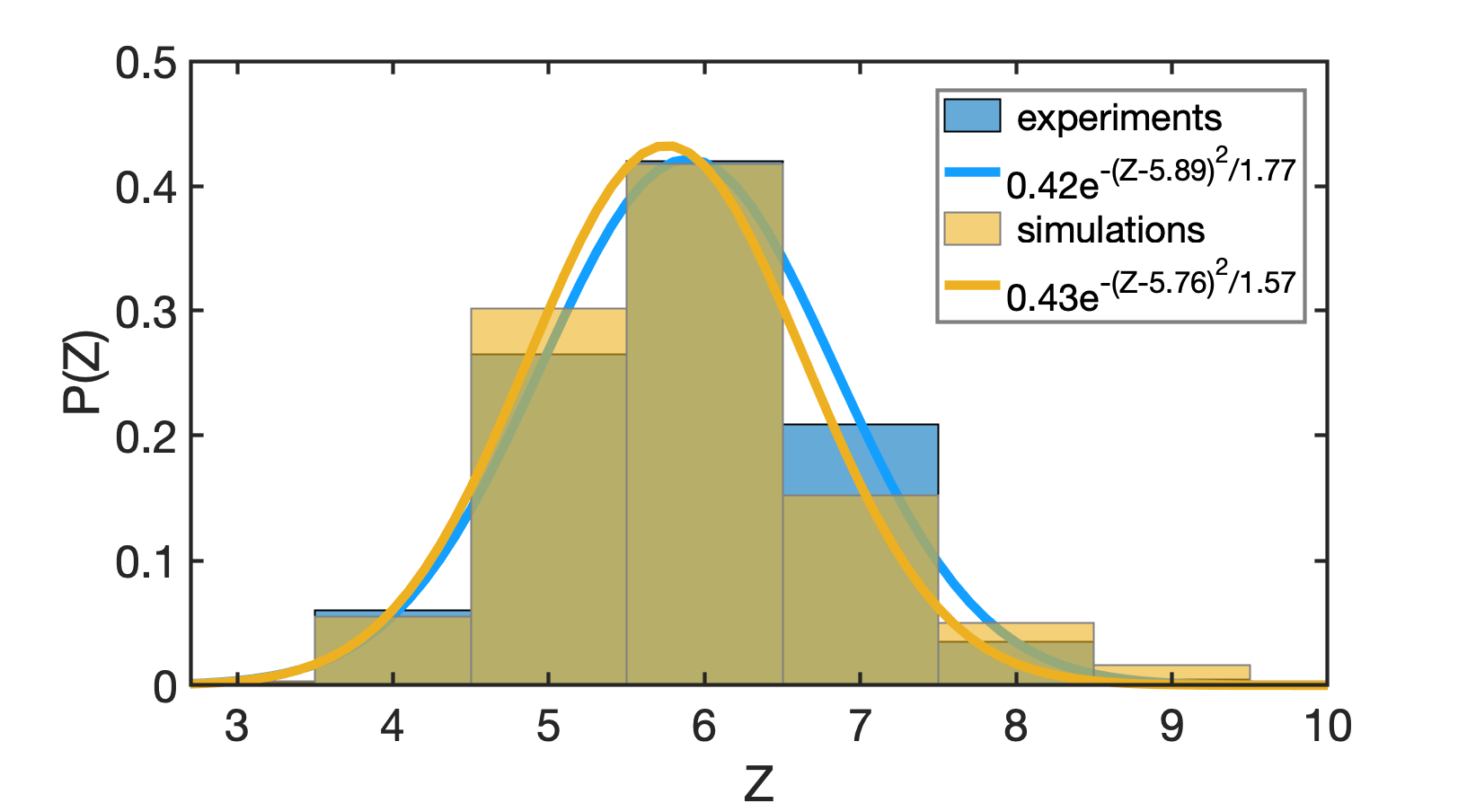} \label{numofneighboringcells} }
		\par\bigskip
		\sidesubfloat[b]{\includegraphics[width=1.0\textwidth] {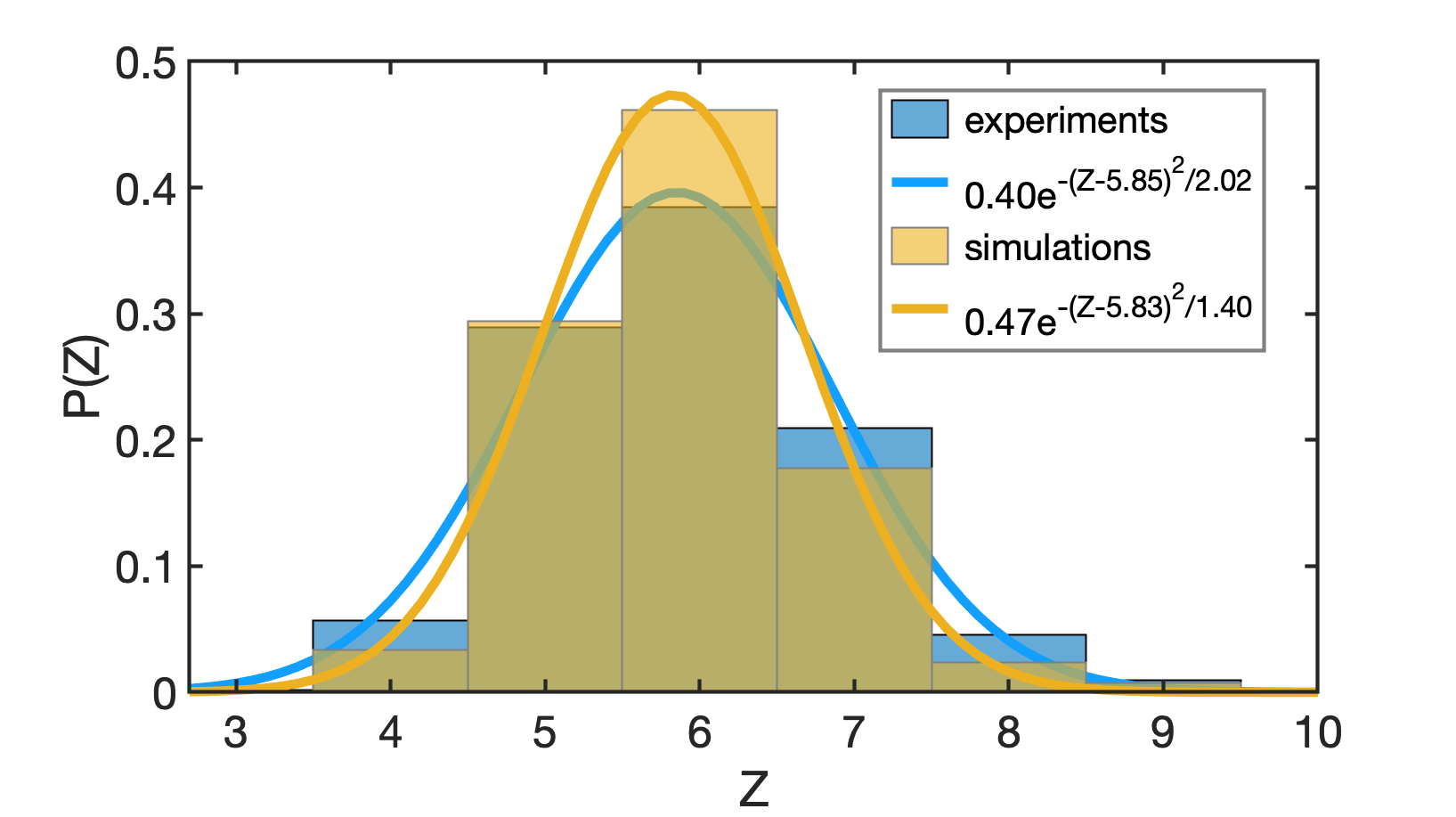} \label{numofneighboring1000} }
\caption{\textbf{Distribution of cell coordination number ($Z$).} (a) The distribution ($P(Z)$) of $Z$ between $120-129$ minutes. (b) Same as (a) except the  time interval $1000-1009$ minutes. The blue (brown) histograms are calculated from experiments (simulations). The solid blue and brown lines are Gaussian fits to the experimental and simulation results, respectively.  }
\label{numofneighboringcells}
\end{figure}


\clearpage
\begin{figure}
{\includegraphics[clip,width=1.05\textwidth]{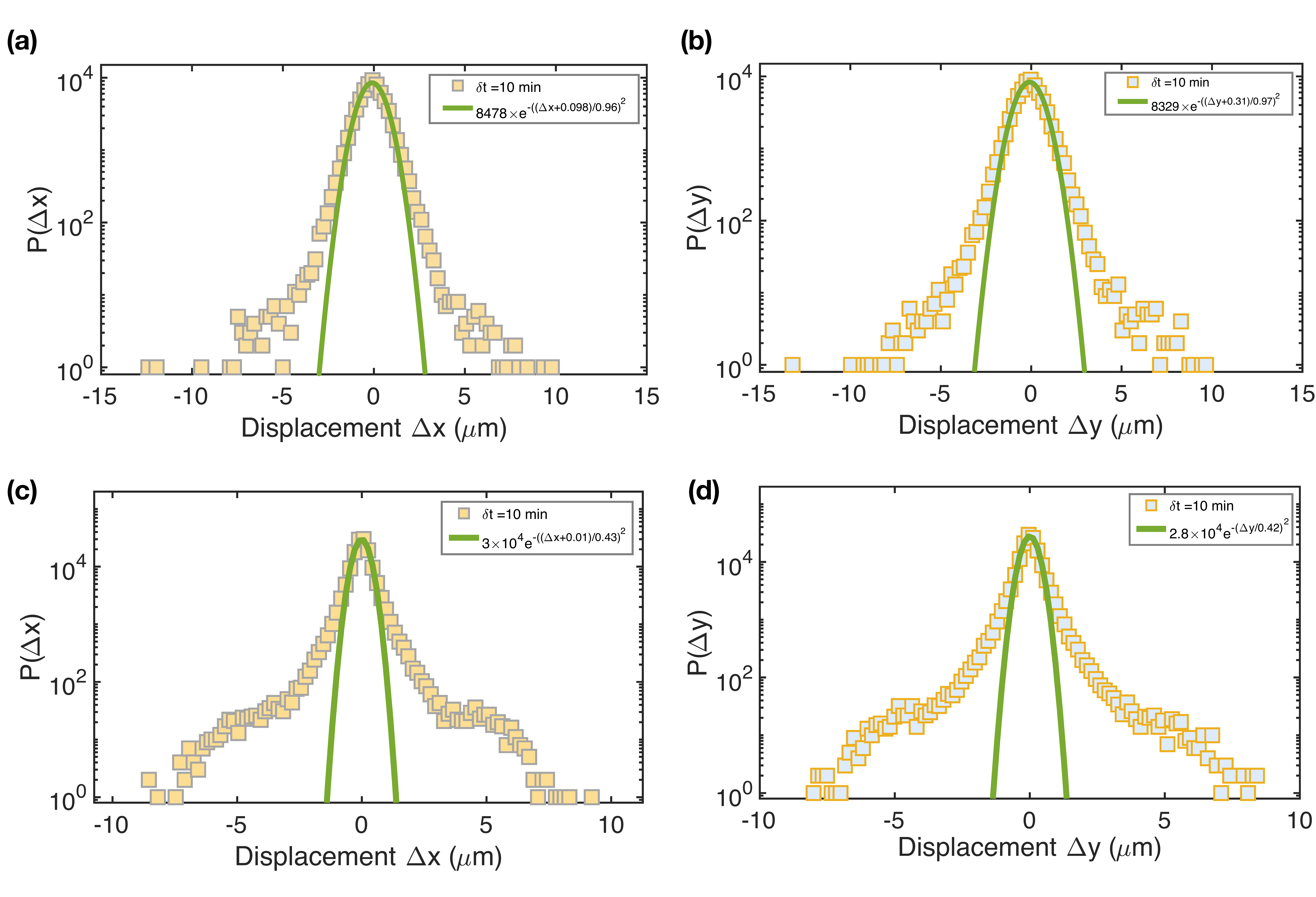}}
\caption{\label{displacemsd} \textbf{Anomalous diffusion.}   (a-b) Distribution (P($\Delta$x), P($\Delta$y)) of cell displacement at the time interval, $\delta t$, of 10 minutes calculated for all the cells over the whole experiment. The green lines show Gaussian distributions.   (c-d) Same as (a-b), except the distributions are obtained from simulations.  }
\label{msddis}
\end{figure}

\clearpage
\begin{figure}
{\includegraphics[clip,width=1.0\textwidth]{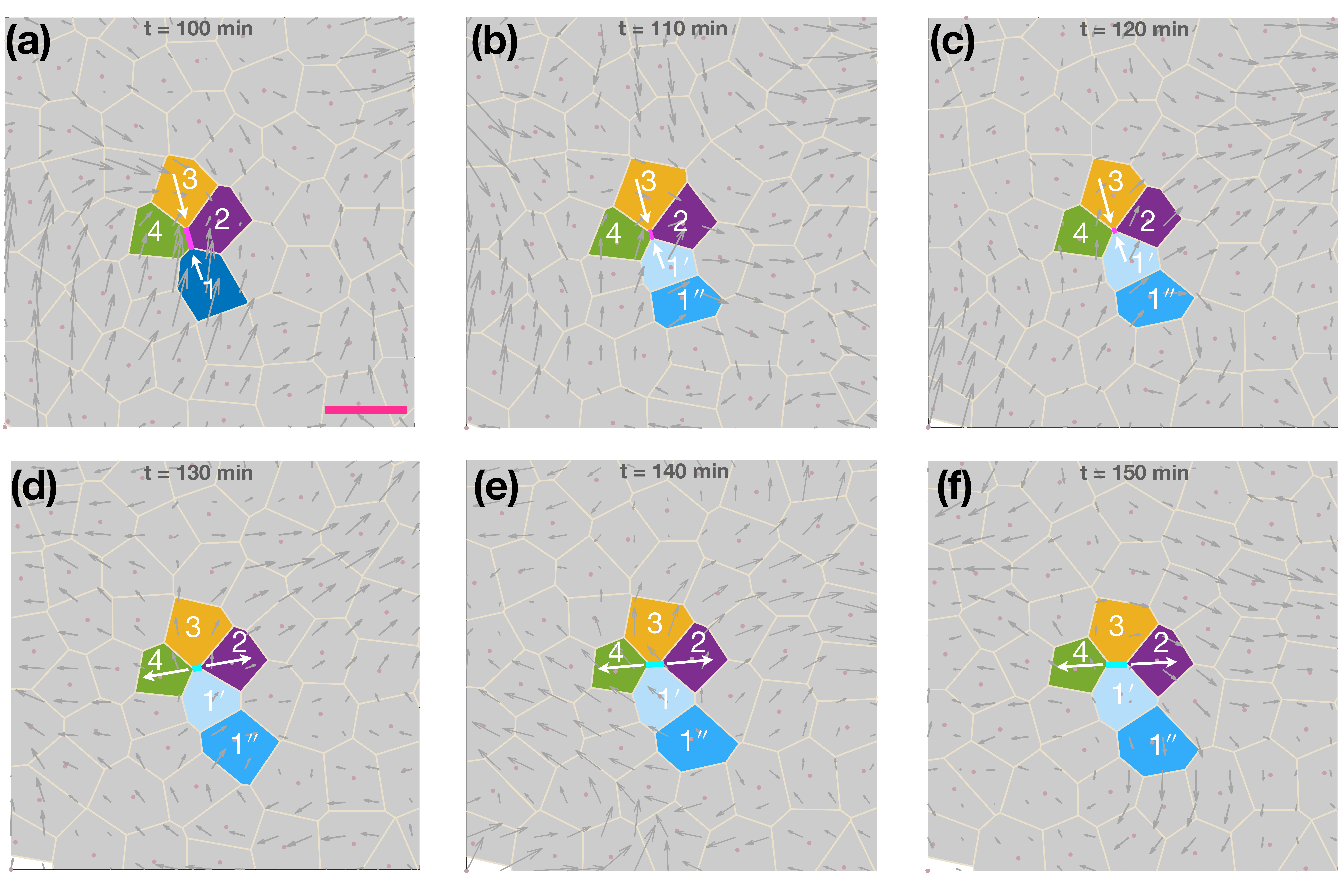}}
\caption{\textbf{T1 topological changes.} Same as Figs.~2e-g in the main text, except  more time frames are included. There is a T1 transition, accompanied by one vortex formation, which is induced by a single cell division (cell $1$ divided into cells $1'$ and $1''$) . The pink bar in (a) corresponds to 20 $\mu m$. The velocity vector field is illustrated by grey arrows.}
\label{clock}
\end{figure}



\clearpage
\begin{figure}
{\includegraphics[clip,width=1.02\textwidth]{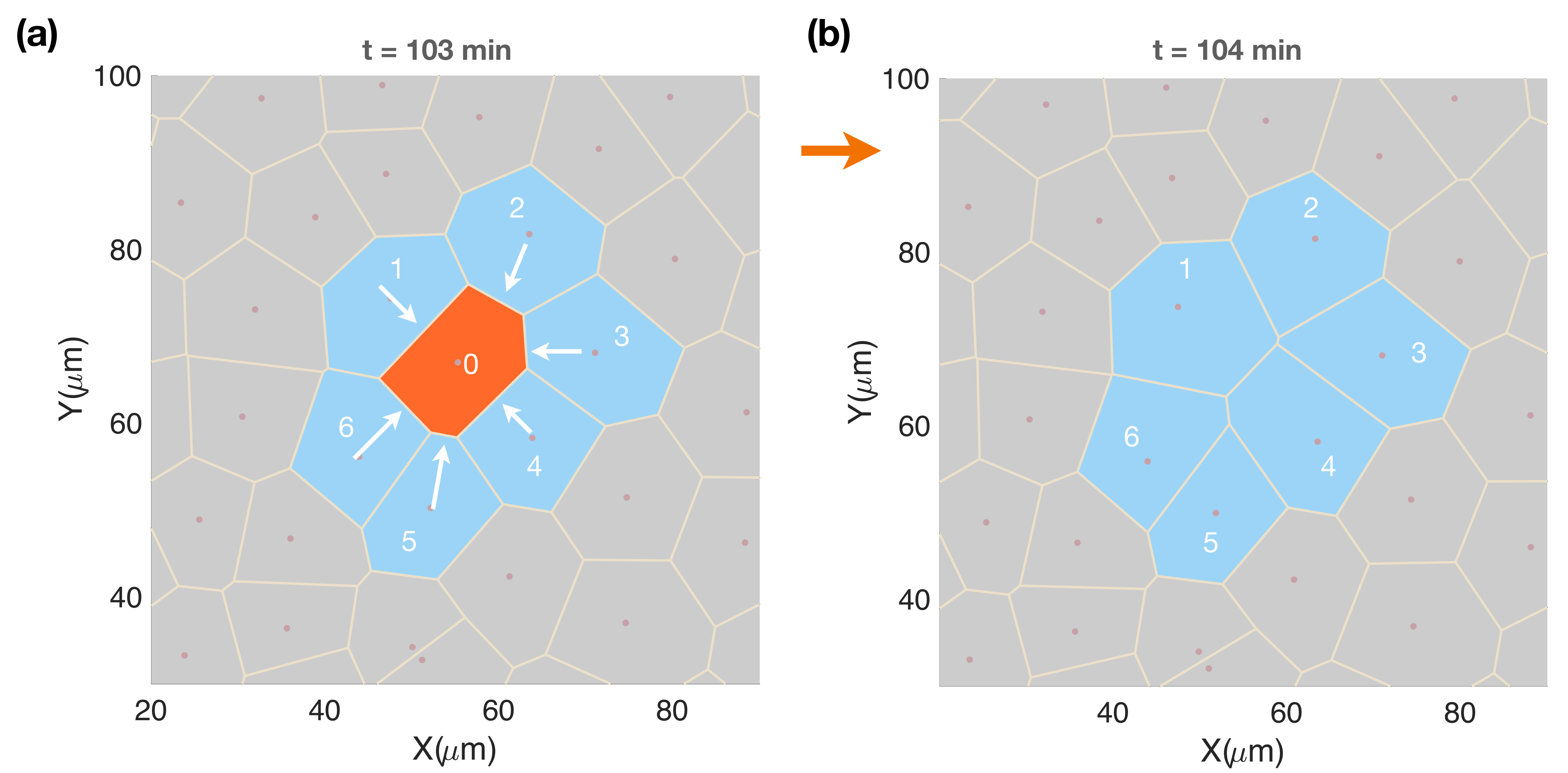}}
\caption{\textbf{Cell apoptosis (extrusion) induces T2 transition.}   Cell $0$ (red), surrounded by cells $1-6$ (blue) at time $t=103$ min (a) is removed from the cell monolayer at time $t=104$ min due to apoptosis/extrusion process, which results in the topological T2 transition (shown in (b)).}
\label{T2transitionfinal}
\end{figure}

\clearpage
\floatsetup[figure]{style=plain,subcapbesideposition=top}
\begin{figure}
{\includegraphics[clip,width=1.05\textwidth]{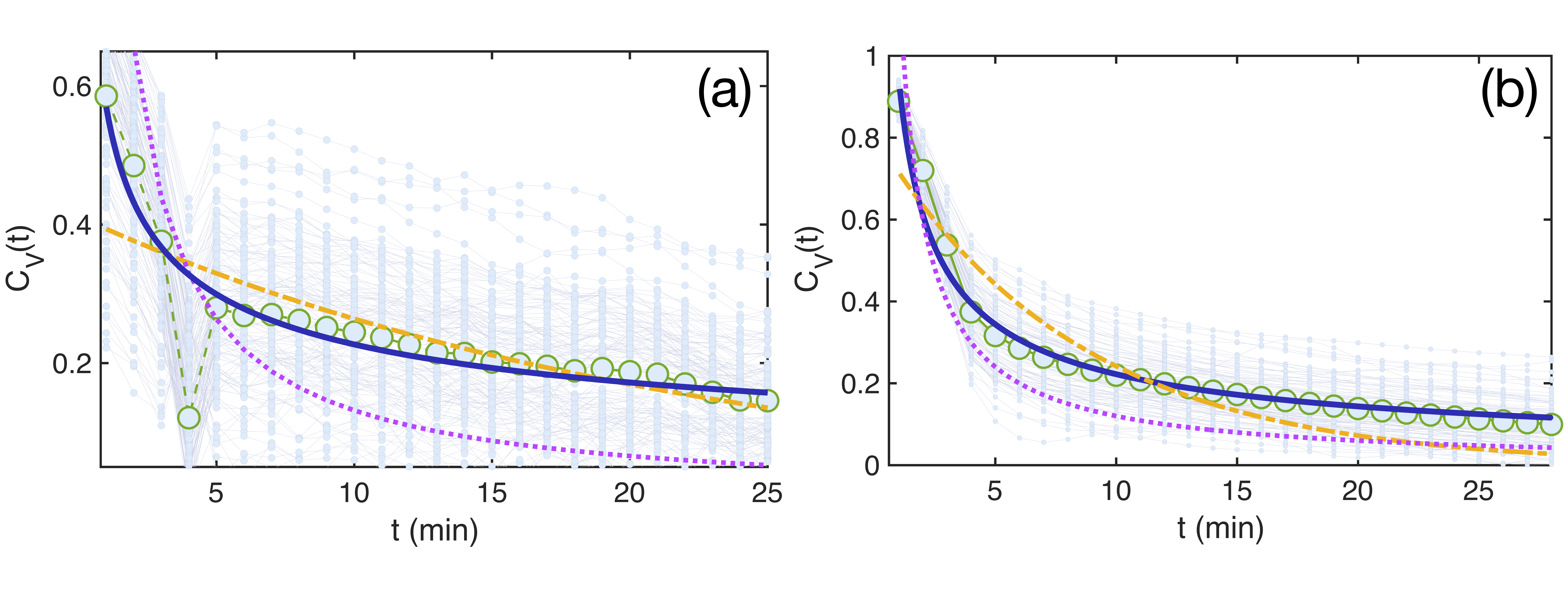}}
\caption{\label{cvtm} \textbf{Velocity autocorrelation:} Same as Figs.~4(c-d) in the main text. (a) Experimental, and (b) simulation data fit using different functions. The blue solid line shows a power-law fit ($t^{-\beta}$ with $\beta \approx 0.4, 0.6$ as listed in Figs.~4(c-d)). The yellow dash-dotted line indicates an exponential fit, and the violet dotted line shows a fit using $t^{-1}$, the expected behavior for 2D liquids. It is clear that the best fit is obtained using $t^{-\beta}$ (with $\beta < 1$). }
\end{figure}

\clearpage
\floatsetup[figure]{style=plain,subcapbesideposition=top}
\begin{figure}
{\includegraphics[clip,width=1.05\textwidth]{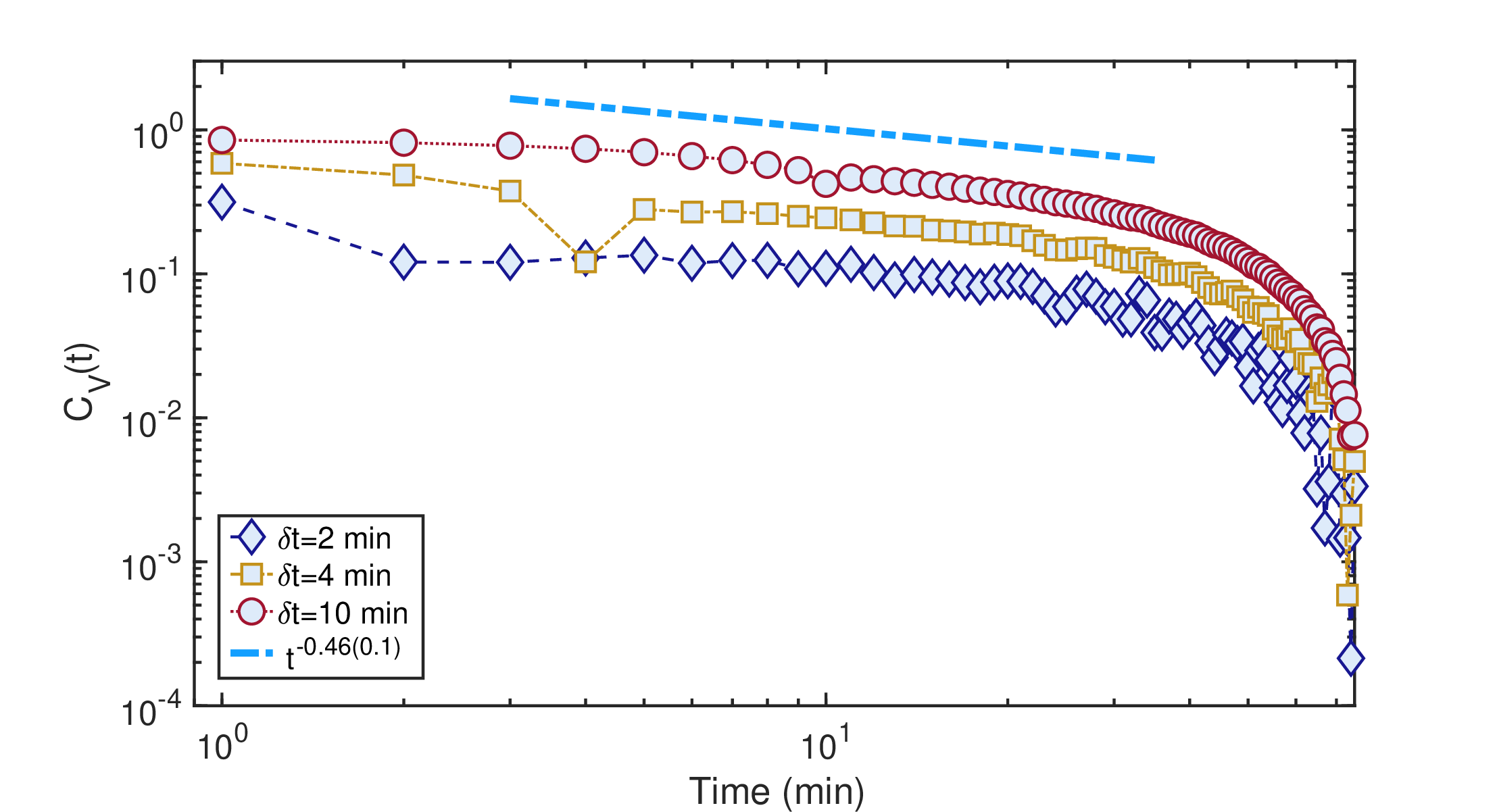}}
\caption{\label{cvtm} \textbf{Velocity autocorrelation at different $\delta t$:} (a)  Same as Fig.~4(c) in the main text. We calculated $C_{v}(t)$ with $\delta t$ = 2, 4, and 10 minutes. The slope of the dashed-dotted line is shown in the box with errors listed in the bracket. Thus, variations in $\delta t$ do not change the emergence of long time tails. }
\end{figure}

\clearpage
\begin{figure}
{\includegraphics[clip,width=1.02\textwidth]{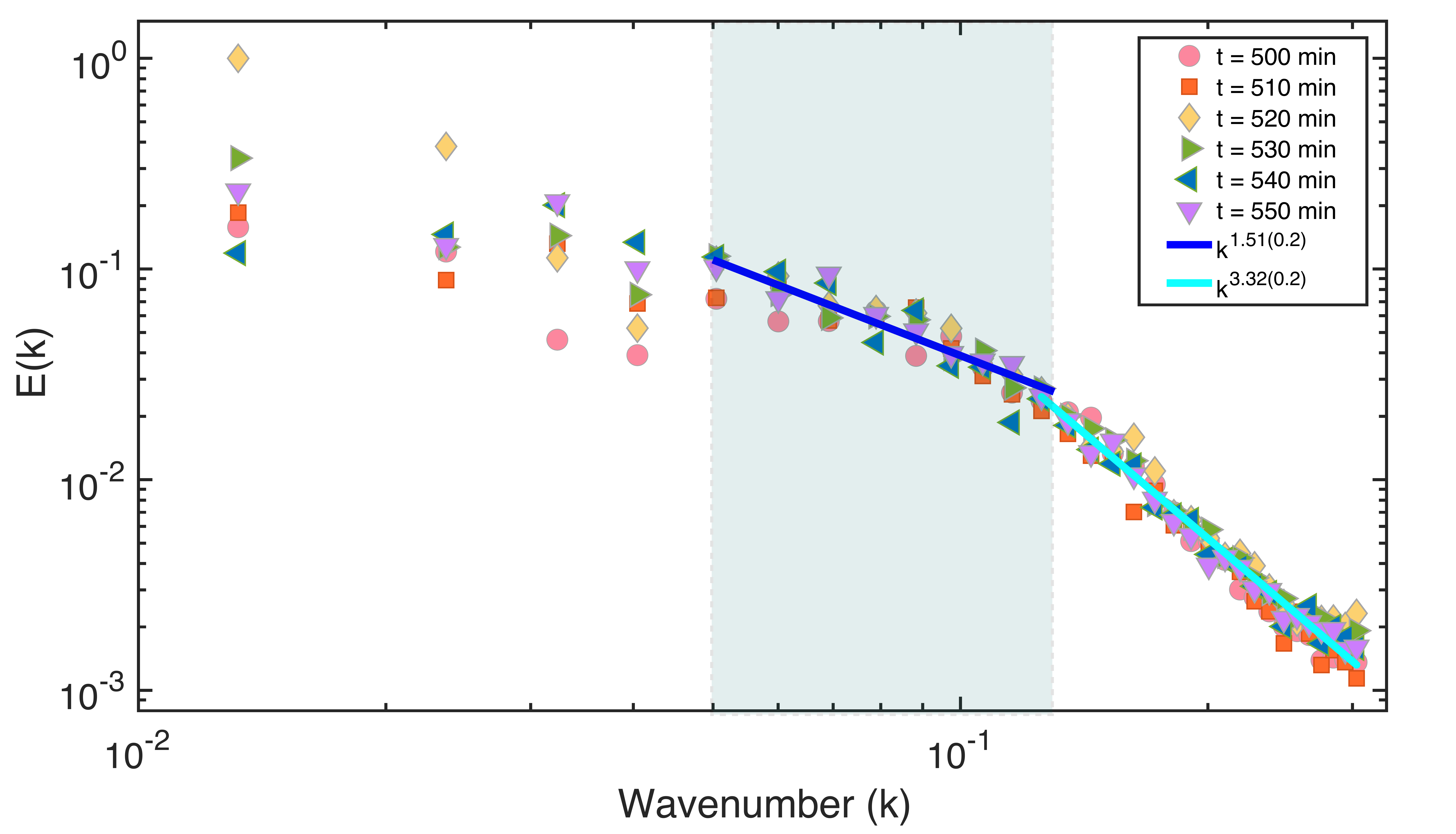}}
\caption{\textbf{The energy spectra E(k) at different times.} Same as Figure 4(e) in the main text, except at different times ($t = 500-550$ minutes). The values of the slope of the solid lines are shown in the upper right corner with errors listed in the bracket. The unit for the wave number ($k$) is $\mu$m$^{-1}$, and  $E(k)$ ($\mu$m$^3$/s$^2$) is scaled by the maximum value.  }
\label{EK500}
\end{figure}

\noindent